\documentclass[usegraphicx,useAMS,usenatbib]{mn2e}
\usepackage{url}
\usepackage{lscape}   
\usepackage{multicol} 
\usepackage{verbatim} 
\usepackage{multirow} 
\usepackage{rotating} 
\usepackage{wasysym}
\usepackage{xspace}

\def \hi{H{\sc i}\xspace}
\def \hii{H{\sc ii}\xspace}
\def \kms{\mbox{km\,s$^{-1}$}\xspace}
\def \Kkms{\mbox{K\,km\,s$^{-1}$}\xspace}
\def \vlsr{V$_{\rm LSR}$\xspace}
\def \arcsec{{$^{\prime\prime}$}\xspace}
\def \cmcube{\mbox{cm$^{-3}$}\xspace}
\def \cmsqr{\mbox{cm$^{-2}$}\xspace}
\def \trms{\mbox{T$_{\rm RMS}$}\xspace}
\def \vrms{\mbox{$v_{\rm RMS}$}\xspace}
\def \tmb{\mbox{T$_{\rm mb}$}\xspace}
\def \ta{\mbox{T$_{\rm A}^{*}$}\xspace}
\def \tsys{\mbox{T$_{\rm SYS}$}\xspace}
\def \nh3{\mbox{NH$_{3}$}\xspace}
\def \C34S{\mbox{C$^{34}$S}\xspace}
\def \13CS{\mbox{$^{13}$CS}\xspace}
\def \tkin{\mbox{$T_{\rm k}$}\xspace}
\def \msun{\mbox{M$_{\odot}$}\xspace}
\def \livedata{\mbox{\sc livedata}\xspace}
\def \gridzilla{\mbox{\sc gridzilla}\xspace}
\def \ASAP{\mbox{\sc asap}\xspace}
\def \miriad{\mbox{\sc miriad}\xspace}
\def \etal{\emph{et al.}\xspace}

\title[7\,mm line survey: W28 field TeV sources.]
{A 7\,mm line survey of the shocked and disrupted molecular gas towards the W28 field TeV gamma-ray sources.}
\author[B. Nicholas et al.]
       {B. P. Nicholas$^{1}$\thanks{E-mail: brent.nicholas@adelaide.edu.au}, G. Rowell$^{1}$, M. G. Burton$^{2}$, A. J. Walsh$^{3}$, Y. Fukui$^{4}$, A. Kawamura$^{4}$\newauthor and N. I. Maxted$^{1}$.\\
$^{1}$School of Chemistry and Physics, Adelaide University, Adelaide, 5005,  Australia\\
$^{2}$School of Physics, University of New South Wales, Sydney, 2052, Australia\\
$^{3}$Centre for Astronomy, School of Engineering and Physical Sciences, James Cook University, Townsville, 4811, Australia\\
$^{4}$Department of Astrophysics, Nagoya University, Furocho, Chikusa-ku, Nagoya, Aichi, 464-8602, Japan}

\begin{document}

\date{\today}


\maketitle

\label{firstpage}

\begin{abstract}

We present 7\,mm Mopra observations of the dense molecular gas towards
the W28 supernova remnant (SNR) field, following a previous 12\,mm
line survey of this region. These observations take advantage of the
7\,mm beam size to probe the dense and disrupted gas in the region at
$\sim$1 arcmin scales. Our observations are focused towards the
north-eastern (NE) HESS J1801-233 and southern HESS J1800-240\,B TeV
gamma-ray sources, with slightly less observations towards HESS J1800-240\,A \&
C. Using the CS\,(1-0) transition we reveal multiple regions of dense
gas, $n_{\rm H_2} \sim10^{5}$\,\cmcube. We report the discovery of
dense gas towards HESS J1800-240\,C, at the site of a 1720\,MHz OH
maser. The NE molecular cloud is known to be disrupted, many 1720\,MHz
OH masers and broad CO line emission are detected at the rim of
W28. Here we reveal this shock interaction region contains generally extended clumpy CS,
as well as clumpy SiO and CH$_3$OH emission with broad line profiles. The FWHM of the
molecular lines extend up to 18\,\kms on the W28 side of the NE
cloud. The detection of SiO towards maser clumps OH C, D, E \& F
provide further evidence of the shocked conditions in the NE cloud.
Several other lines associated with star
formation are also detected towards the southern source, notably the
energetic \hii complex G5.89-0.39. The spatial match of dense gas with the TeV emission 
further supports the CR origin for the gamma-rays. We estimate the mass of
several extended dense clouds within the field and predict the TeV
flux from the dense cloud components. The predicted fluxes are on the
order of $10^{-14}$--$10^{-13}$\,ph\,\cmsqr\,s$^{-1}$, which should be
detectable and possibly resolved by a future TeV instrument, such as
the Cherenkov Telescope Array.

\end{abstract}


\section{Introduction}

W28 is a striking example of TeV (10$^{12}$~eV) gamma-ray emission
spatially overlapping with molecular gas~\citep{hess_w28}. W28 is part
of the increasing list of sources with spatial overlap between TeV or
GeV (10$^{9}$~eV) gamma-ray emission and molecular gas
(e.g. HESS~J1745-290/SNR~G359.1-0.5~\citep{hess_1745},
HESS~J1714-385/CTB~37A~\citep{hess_ctb37a},
HESS~J1923+141/SNR~G49.2-0.7~\citep{feinstein},
IC~443~\citep{ic443_magic,ic443_veritas}), in addition to the central
molecular zone (CMZ) towards the Galactic centre
region~\citep{hess_CMZ}.

W28 is a mixed morphology old age ($>10^{4}$\,yr;~\citealt{kaspi})
supernova remnant (SNR) with dimensions of 50$^\prime \times
45^\prime$. W28 is estimated to be at a distance of 1.2 to 3.3\,kpc
(e.g.~\citet{goudis,lozinskaya,motogi}).  It has been shown to exhibit
non-thermal radio emission~\citep{dubner}, thermal X-ray emission
\citep{rho2002}, and, more recently gamma-ray emission at
TeV~\citep{hess_w28} and GeV~\citep{fermi_w28, agile} energies.

Several CO surveys in the (1-0), (2-1) and (3-2) lines reveal massive
molecular clouds to the north-east (NE) and to the south (S) of the
SNR \citep{arikawa,reach,torres,hess_w28,nanten21}. Most of the CO
emission appears centred at a local standard of rest velocity (\vlsr)
similar to that inferred for W28 \vlsr$\sim$7\,\kms (corresponding to
a distance $\sim 2$\,kpc) based on \hi studies \citep{velazquez}.

The molecular clouds NE of W28 are known to exhibit broad CO line
emission. It has been argued that W28 has disrupted much of this gas,
giving rise to its broad velocity distribution~\citep{arikawa, reach,
  torres}. The most likely mechanism for this type of disruption would
be the SNR shock interacting with the molecular gas, as is the case
for other sources
e.g. HESS~J1745-290/SNR~G359.1-0.5~\citep{hess_1745},
HESS~J1714-385/CTB~37A~\citep{hess_ctb37a},
HESS~J1923+141/SNR~G49.2-0.7~\citep{feinstein} and
IC~443~\citep{ic443_magic,ic443_veritas}. All of these sources
similarly display 1720\,MHz OH masers and are mature SNRs (age
$>10^{4}$\,yr). Our \nh3 observations~\citep{me} revealed that the W28
SNR shock has disrupted the dense core of the NE cloud, and strong
\nh3(3,3) and (6,6) detections with broad line widths (FWHM
$>10$\,\kms), suggest that this region is warm and turbulent.  The
shocked NE cloud also provides an opportunity to study the diffusion
and propagation of CRs into molecular clouds. Additionally, the
effects of a SNR shock propagating through a dense molecular cloud can
be studied and search for sites of star formation triggered by the
passing shock.

The southern clouds harbour sites of high mass star formation,
containing multiple \hii regions, G6.225-0.569,
G6.1-0.6~\citep{lockman,kuchar} and the ultracompact \hii region
(UC\hii) G5.89-0.39, (see e.g~\citet{harvey, kimkoo2001}). A recently
detected 1720\,MHz OH maser towards a candidate SNR
G5.71-0.08~\citep{brogan} may also suggest that there is another SNR
shock with molecular gas interaction occurring in one of the southern
clouds. A key question is whether the W28 SNR has disrupted the
southern clouds and is responsible for the TeV emission, or whether
other high energy processes are at play. The star formation activity
present in the southern clouds may play a part in the production of
TeV emission, as theory into protostellar particle acceleration may
suggest~\citep{araudo}.

The the TeV gamma-ray emission detected towards the NE cloud has strong
evidence supporting the hadronic emission mechanism. In order to model
leptonic scenarios, uncomfortably low magnetic fields and gas
densities are required
~\citep{yamazaki2006,fujita2009,gabici2010,fermi_w28}. These
requirements contradict with the observed molecular gas densities
$n_{\rm H_2} > 10^{4}$\,\cmcube~\citep{me} and the magnetic field
enhancements seen in dense molecular clouds~\citep{crutcher}.  The key
question relates to the origin of the TeV particles. Provided the
hadronic emission scenario, a population of high energy particles are
required to interact with the dense clouds. For the NE cloud, the SNR
shock is the obvious source of accelerated particles. However, the
same can not be conclusively said about the southern sources. CRs
accelerated by W28 in the past may have propagated south to illuminate
the southern clouds. Additionally, multiple \hii regions and sites of
star formation, including the energetic UC\hii complex G5.89-0.39,
could assist in CR acceleration towards the southern TeV sources.

Since the common and abundant gas tracer, CO, with a critical density
of 10$^{2}$\,\cmcube, rapidly becomes optically thick towards clumps
and cores, probing the molecular cloud density profile can be
impaired. Ideal tracers of dense gas are those with a lower abundance
(10$^{-5}\times$CO) and higher critical density
(10$^{4}$-$10^{5}$\,\cmcube). Using the Mopra telescope, we \citep{me}
conducted broad scale ($\sim1.5^{\circ}$\,square) observations of the
W28 field in a 12\,mm line survey. Taking advantage of the 23\,GHz
\nh3 inversion lines, the cold dense interiors of the molecular clouds
were probed and revealed dense gas spatially consistent with both the
CO gas and TeV emission.

In order to further probe the structure and details of the W28 field
molecular clouds, we have continued to use the Mopra radio telescope
in a 7\,mm line survey to observe the CS\,(1-0) line, as well as other
lines including SiO\,(1-0), Class I CH$_{3}$OH masers and
cyanopolyynes HC$_{n}$N ($n=3,5,7$), to trace the dense gas, the
presence of shocks, outflows, and disrupted gas, and sites of
high-mass star formation.

The 43\,GHz CS\,(1-0) line is a useful warm, dense gas tracer, and, as
CS also exists in various isotopologue forms, the optical depth and
column density can be constrained. The higher frequency of 7\,mm lines
compared to our previous 12\,mm observations, results in a smaller
beam FWHM ($\sim1$ arcminute). This makes CS an ideal follow-up tracer
to determine the properties of the dense and disrupted gas toward the
molecular clouds in the W28 region.

\section{Mopra Observations and Data Reduction}
\label{sec:obs}

Observations were performed with the Mopra radio telescope in April of
2009 and March of 2010 and utilised the UNSW Mopra wide-band
spectrometer (MOPS) in zoom mode. Mopra is a 22\,m single-dish radio
telescope located $\sim$450\,km north-west of Sydney, Australia
($31^\circ 16^\prime 04^{\prime\prime}$\,S, $149^\circ 05^\prime
59^{\prime\prime}$\,E, 866m a.s.l.). The 7\,mm receiver operates in
the 30-50\,GHz range and, when coupled with MOPS, allows an
instantaneous 8\,GHz bandwidth. This gives Mopra the ability to cover
40\% of the 7\,mm band and simultaneously observe many spectral
lines. The zoom mode of MOPS allows observations in up to 16 windows
simultaneously, where each window is 137.5\,MHz wide and contains 4096
channels in each of two polarisations. At 7\,mm this gives MOPS an
effective bandwidth of $\sim$1000\,\kms with resolution of
$\sim$0.2\,\kms. Across the whole 7\,mm band, the beam FWHM varies
from 1.37$^{\prime}$ (31\,GHz) to 0.99$^{\prime}$
(49\,GHz)~\citep{mopra_beam}. Table~\ref{tab:lines} lists the lines
which MOPS was tuned to receive.

\begin{table}
\centering
\caption{Molecular lines and their corresponding rest frequencies
  which MOPS was tuned to receive. The final two columns indicate
  whether the line was detected in our mapping or deep pointing
  observations.\label{tab:lines}} \normalsize
\begin{tabular}{llcc}
\hline
Molecular Line & Frequency & Detected & Detected \\
\multicolumn{1}{c}{Name} & \multicolumn{1}{c}{(MHz)} & Map & Deep Spectra\\
\hline
$^{30}$SiO\,(1-0, $v=0$) & 42373.365 & -- & --\\
SiO\,(1-0, $v=3$) & 42519.373 & Yes & --\\
SiO\,(1-0, $v=2$) & 42820.582 & Yes & --\\
$^{29}$SiO\,(1-0, $v=0$) & 42879.922 & -- & --\\
SiO\,(1-0, $v=1$) & 43122.079 & Yes & --\\
SiO\,(1-0, $v=0$) & 43423.864 & Yes & Yes\\
CH$_{3}$OH-I & 44069.476 & Yes & Yes\\
HC$_{7}$N\,(40-39) & 45119.064 & -- & --\\
HC$_{5}$N\,(17-16) & 45264.75 & Yes & Yes\\
HC$_{3}$N\,(5-4) & 45488.839 & Yes & Yes\\
\13CS\,(1-0) & 46247.58 & Yes & Yes\\
HC$_{5}$N\,(16-15) & 47927.275 & Yes & Yes\\
\C34S\,(1-0) & 48206.946 & Yes & Yes\\
OCS\,(4-3) & 48651.6043 & -- & --\\
CS\,(1-0) & 48990.957 & Yes & Yes\\
\hline
\end{tabular}
\end{table}

`On-the-fly' (OTF) mapping observations were based on our earlier
12\,mm mapping of the W28 region. Although we mapped all four TeV
sources, deeper observations towards the NE shocked cloud (HESS
J1801-233) and southern cloud (HESS J1800-240\,B) were obtained than
toward HESS J1800-240\,A \& C. The observations were recorded
alternating the scanning direction from RA to Dec. to reduce noise
levels and to eliminate artificial stripes that can be introduced when
only one scanning direction is used. Collectively, over both
observation runs we obtained 9 passes on the NE cloud (HESS
J1801-233), 9 passes towards the southern cloud (HESS J1800-240\,B)
and 3 passes toward each of the HESS J1800-240\,A and HESS
J1800-240\,C regions.

Data were reduced using the standard ATNF packages {\livedata},
{\gridzilla}, {\ASAP} and {\miriad}.\footnote{See
  \url{http://www.atnf.csiro.au/computing/software/} for more
  information on these data reduction packages.} For mapping data,
{\livedata} was used to perform a bandpass calibration for each row
using the preceding off-scan as a reference, and applied a linear fit
to the baseline. {\gridzilla} re-gridded and combined all data from
all mapping scans into a single data cube, with pixels ($\Delta
x,\Delta y,\Delta z$)=(15\arcsec, 15\arcsec, 0.21\,\kms). The mapping
data were also weighted according to the relevant \tsys,
Gaussian-smoothed based on the Mopra beam (1.2$^\prime$ FWHM and
3$^\prime$ cut-off radius), and pixel masked to remove noisy edge
pixels. Data cubes were then converted into an {\ASAP} scantable and
had a $7^{{\rm th}}$ order polynomial function fit subtracted from
each\ pixel in the cube before returning the final output fits
files. Analysis of position-switched deep pointings employed {\ASAP}
with time-averaging, weighting by the relevant \tsys, and baseline
subtraction using a linear fit after masking the 15 channels at each
bandpass edge.

In both mapping and position-switched data, the antenna temperature
\ta (corrected for atmospheric attenuation and rearward loss) was
converted to the main beam brightness temperature \tmb (K), such that
\tmb = \ta /$\eta_{\rm mb}$ where $\eta_{\rm mb}$ is the Mopra main beam
efficiency. Each line was corrected by the relevant $\eta_{\rm mb}$
following~\citet{mopra_beam}.

\begin{figure*}
\includegraphics[width=\textwidth]{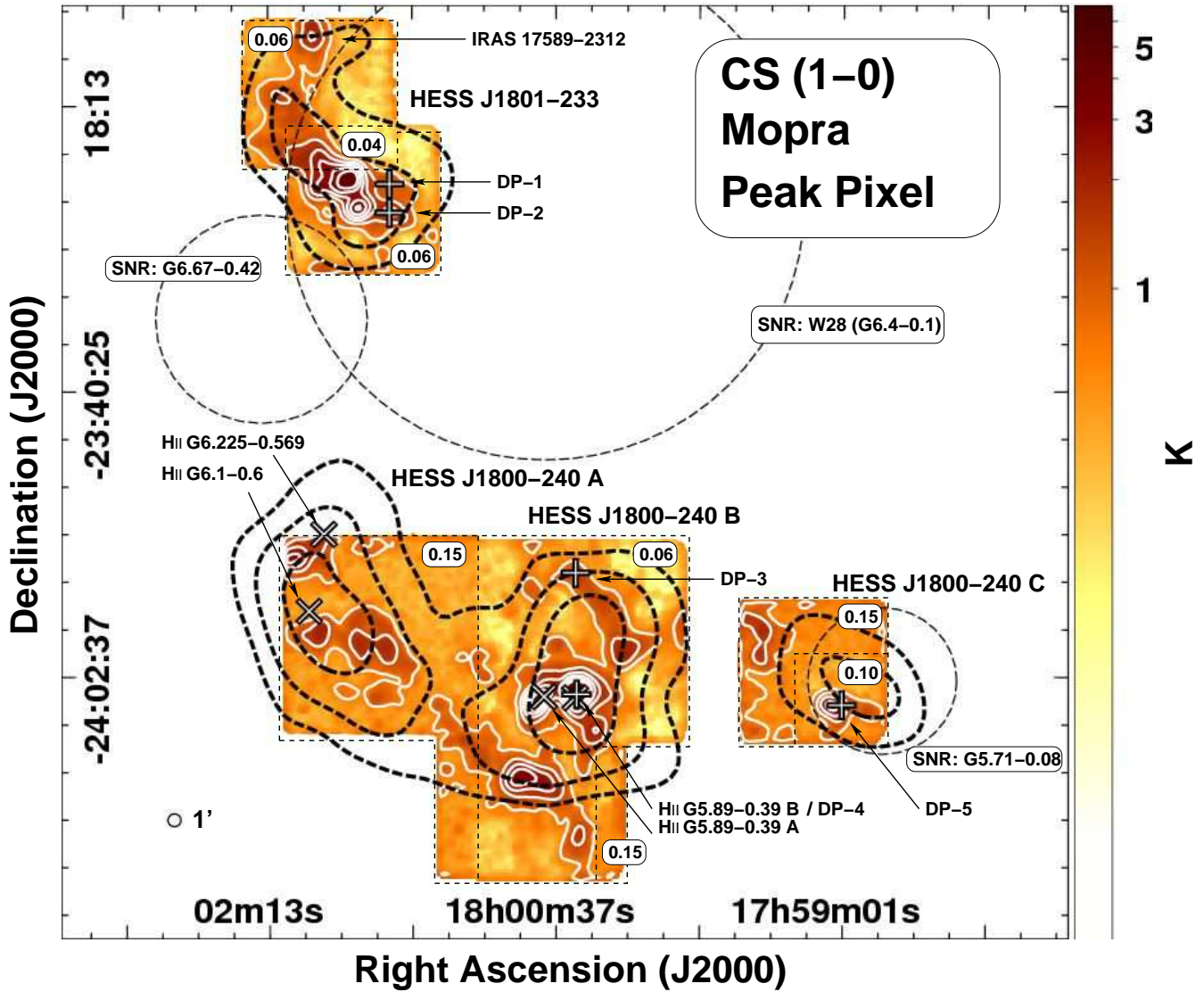}
\caption{CS\,(1-0) peak pixel map of the W28 region (log scale image
  with white linear contours, min. contour $\sim10$\,\trms) with
  black/white crosses $+$ to indicate the locations of the
  position-switched deep pointings (DP-1 to DP-5). Thick black dashed
  contours are the H.E.S.S. TeV emission (4$\sigma$, 5$\sigma$,
  6$\sigma$ levels) revealing HESS J1801-233 and the different
  components of HESS J1800-240. Boundaries of catalogued SNRs are
  indicated by dashed circles, and locations of \hii regions are
  indicated by black/white crosses $\times$. Black dashed boxes are
  used to separate regions based on their exposure. The numbers
  presented in the white boxes are the mean main beam \trms (units of
  K) achieved across the 42-49\,GHz band within that boundary. The 1
  arcmin beam FWHM is also shown in the lower left corner.}
\label{fig:exposure}
\end{figure*}

\begin{figure*}
\includegraphics[width=\textwidth]{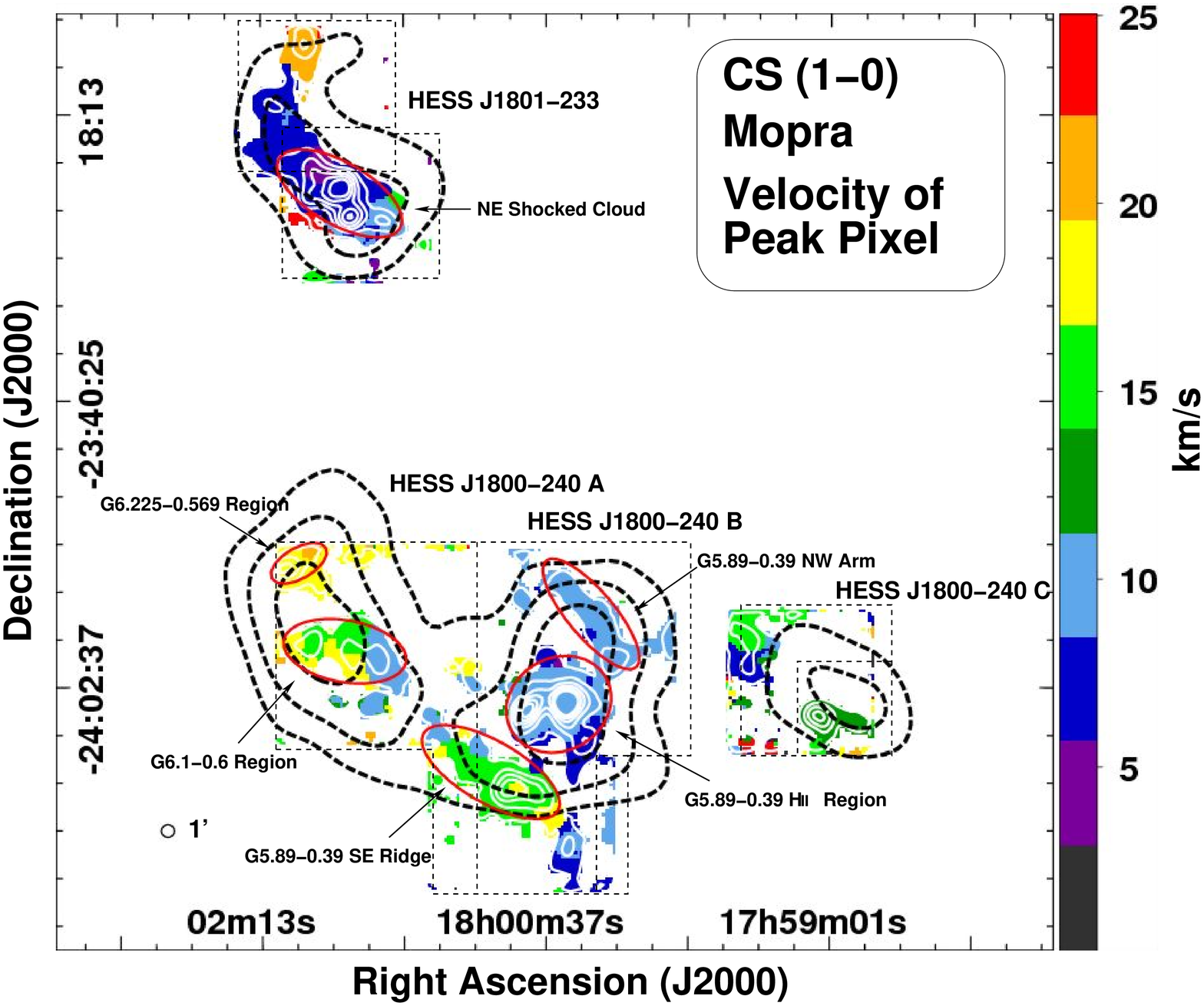}
\caption{Velocity of the peak pixels, with the same black and white
  contours as shown in Figure~\ref{fig:exposure}. Red solid ellipses are the
  regions where spectra were averaged in order to extract gas
  parameters as an extended source.}
\label{fig:whole_velopix}
\end{figure*}

Due to the uneven mapping exposure across the W28 field, the RMS error
in \tmb, \trms (K), varies. On top of the exposure dependence, there
is also a frequency dependence on \trms across the 42-49\,GHz band.
For this reason in Figure~\ref{fig:exposure} we present a CS\,(1-0)
peak pixel map of the W28 field with dashed boxes outlining the
exposure boundaries. The numbers presented in the white boxes are the
averaged \trms achieved (across the 42-49\,GHz band) within that
boundary. We also show in Figure~\ref{fig:whole_velopix} the velocity
of the peak pixels to show the general velocity locations and
distribution of the various clumps of CS\,(1-0) gas. We show the same
contours on both Figures~\ref{fig:exposure}
and~\ref{fig:whole_velopix} for guidance.

For the detected lines, images of velocity-integrated intensity,
position velocity (PV) and velocity dispersion (\vrms) have been
produced where possible (although not all are
shown). Velocity-integrated intensity images (\miriad 0$^{\rm th}$
moment) are integrated over a velocity range determined by the
observed line width. This was chosen to encompass the bulk of the
emission from the region of interest.  Minimum contour levels were set
based on the integrated \trms of the emission. The integrated \trms of
each image was determined by creating additional \miriad 0$^{\rm th}$
moment maps in a velocity space either side of the velocity range of
interest, ensuring the same number of channels were used. These
additional moment maps were used to create pixel distribution
histograms which were Gaussian fitted. In this way we estimated the
integrated \trms from the Gaussian fit results. On all images, the
minimum contour levels are mentioned in terms of the raw value as well
as \trms. Generally the minimum accepted contour level on integrated
maps is 2\,\trms.  PV plots were created by re-ordering the data cubes
axes and Hanning smoothing the velocity axis (width $\sim3$\,\kms) to
improve image quality. The PV plots show the peak pixel along the
declination axis for illustrative purposes.  Intensity weighted
velocity dispersion (\miriad 2 moment), \vrms, maps were calculated
for pixels above a reasonable threshold.

\begin{table*}
\centering
\caption{Calculated gas parameters from CS\,(1-0), \C34S\,(1-0) and
  \13CS\,(1-0) isotopologue ratios from deep pointing observations
  DP-1 to DP-5 (Figure~\ref{fig:exposure}). Columns from left to right
  are: region name, CS\,(1-0) peak \tmb temperature, CS\,(1-0) LSR
  velocity, CS\,(1-0) FWHM, CS\,(1-0) optical depth $\tau$, assumed
  kinetic temperature \tkin, beam corrected total column density
  $N$[CS], hydrogen column density $N$[H$_{2}$], mass, molecular
  hydrogen number density $n_{\rm H_{2}}$ and virial mass range
  $M_{\rm vir}$. The virial mass range is the lower and upper bounds
  when considering a constant, Gaussian and $r^{-2}$ density
  profile. \label{tab:massdens}} \small
\begin{tabular}{lcccccccccc}

\hline
Core /	                &	\tmb	&	\vlsr	&	FWHM	&	$\tau$	&	\tkin$^{*}$	&	$N$[CS]	                        &	$N$[H$_{2}$]	              	&	Mass	               &	n$_{\rm H_{2}}$	                &	Virial Mass			\\
Region	                &	[$K$]	&	[\kms]	&	[\kms]	&		&	[$K$]	                &	[$\times10^{14}$\,\cmsqr]	&	[$\times10^{23}$\,\cmsqr]	&	[M$_{\odot}$]	       &	[$\times10^{5}$\,\cmcube]	&	[$\times10^{2}$\,M$_{\odot}$]	\\
\hline																								
\multicolumn{11}{c}{-- Compact Source --}\\
DP-1	                &	1.1	&	13.1	&	18.1	&	1.0	&	46	                &	16.5	                        &	4.1	                        &	790       &	6.9	                        &	133	--	279 	\\
DP-2	                &	2.4	&	7.4	&	11.2	&	1.6	&	46	                &	28.8	                        &	7.2	                        &	1400      &	12.0	                        &	23	--	81	\\
DP-3$^{\dagger}$	&	1.4	&	9.0	&	2.2	&	--	&	20	                &	0.1	                        &	2.2$\times10^{-2}$                       &	77	  &	0.5	                        &	1.2	--	4.3	\\
DP-4            	&	9.5	&	9.3	&	4.0	&	2.8	&	33	                &	61.1	                        &	15.3	                        &	2900      &	25.4	                        &	3.4	--	11.9	\\
DP-5$^{\ddagger}$	&	2.8	&	13.0	&	1.1	&	2.1	&	20	                &	4.0	                        &	1.0	                        &	190       &	1.7	                        &	0.3	--	0.9	\\
\hline
\multicolumn{8}{l}{\small \underline{Compact-source mass/density scaling factors vs. radius $R$ (pc)}}\\
\multicolumn{3}{l}{\small $R$ (pc)}  & \multicolumn{5}{l}{\small 0.10   0.15   0.20   0.25   0.30   0.35   0.40}\\[-0.5mm]
\multicolumn{3}{l}{\small Mass $M(R)$/$M$(0.2\,pc)} & \multicolumn{5}{l}{\small 0.89   0.93   1.00   1.09   0.21   1.36   1.54}\\[-0.5mm]
\multicolumn{3}{l}{\small Density  $n_{\rm \tiny H_{2}}$($R$)/$n_{\rm \tiny H_{2}}$(0.2\,pc)}& \multicolumn{5}{l}{\small 7.09   2.21   1.00  0.56  0.36   0.25   0.19}\\
\hline
\multicolumn{5}{l}{\scriptsize $^{*}$ Temperatures from NH$_{3}$ observations~\citep{me}.}\\[-1mm]
\multicolumn{5}{l}{\scriptsize $^{\dagger}$ Using only CS\,(1-0) and assuming optically thin emission.}\\[-1mm]
\multicolumn{5}{l}{\scriptsize $^{\ddagger}$ Using only CS\,(1-0) and \C34S\,(1-0).}\\
\hline
\end{tabular}
\end{table*}

\begin{table*}
\centering
\caption{Calculated gas parameters from CS\,(1-0), \C34S\,(1-0) and
  \13CS\,(1-0) isotopologue ratios for extended sources, using
  averaged spectra from mapping observations. Columns from left to
  right are: region name, peak \tmb temperature, CS\,(1-0) line FWHM,
  optical depth $\tau$, assumed kinetic temperature \tkin, beam
  corrected total column density $N$[CS], hydrogen column density
  $N$[H$_{2}$], mass and molecular hydrogen number density $n_{\rm
    H_{2}}$. We treat the NE shocked cloud mass as an upper limit
  resulting from uncertainties in the molecular abundance
  ratio.\label{tab:extdmassdens}} \normalsize
\begin{tabular}{lcccccccc}

\hline
Region	&	\tmb	&	FWHM	&	$\tau^{*}$	&	\tkin$^{\dagger}$	&	$N$[CS]	        &	$N$[H$_{2}$]	     &           Mass	        &	n$_{\rm H_{2}}$	                \\
	&	[$K$]	&	[\kms]	&		&	[$K$]	        & [$\times10^{14}$\,\cmsqr]	&  [$\times10^{23}$\,\cmsqr] &   [M$_{\odot}$]	&	[$\times10^{4}$\,\cmcube]	\\
\hline
\multicolumn{9}{c}{-- Extended Source --}\\
NE Shocked Cloud$^{a}$	&	1.8	&	7.3	&	2.3	&	35	&	17.8	&	4.4	&	$<$56000	&	6.1	\\
G5.89-0.39 \hii Region$^{b}$	&	2.2	&	3.8	&	3.8	&	25	&	11.6	&	2.9	&	54000	&	2.7	\\
G5.89-0.39 NW Arm$^{c}$	&	0.8	&	3.5	&	--	&	20	&	1.4	&	0.3	&	 4200	&	0.6	\\
G5.89-0.39 SE Ridge$^{d}$	&	0.7	&	3.1	&	--	&	16	&	2.7	&	0.7	&	12000	&	0.9	\\
G6.1-0.6 Region$^{e}$	&	0.7	&	5.1	&	--	&	25	&	1.8	&	0.5	&	6500	&	0.6	\\
G6.225-0.569 Region$^{f}$	&	1.5	&	3.4	&	--	&	21	&	1.8	&	0.5	&	1600	&	1.2	\\
\hline
\multicolumn{9}{l}{\scriptsize $^{*}$ Optical depths are calculated where possible, if no number given then emission is assumed to be optically thin.}\\[-1mm]
\multicolumn{9}{l}{\scriptsize $^{\dagger}$ Temperatures from \nh3 observations~\citep{me}}\\[-1mm]
\multicolumn{9}{l}{\scriptsize $^{a}$ For ellipse $7.2\times2.8$\,pc diam.; pos. angle $-30^{\circ}$; centred on RA 18:01:46.7 Dec -23:24:21.9 shown in Figure~\ref{fig:whole_velopix}}\\[-1mm]
\multicolumn{9}{l}{\scriptsize $^{b}$ For ellipse $4.8\times4.2$\,pc diam.; pos. angle $+30^{\circ}$; centred on RA 18:00:32.9 Dec -24:03:59.2 shown in Figure~\ref{fig:whole_velopix}}\\[-1mm]
\multicolumn{9}{l}{\scriptsize $^{c}$ For ellipse $6.3\times2.1$\,pc diam.; pos. angle $-50^{\circ}$; centred on RA 18:00:22.0 Dec -23:56:59.2 shown in Figure~\ref{fig:whole_velopix}}\\[-1mm]
\multicolumn{9}{l}{\scriptsize $^{d}$ For ellipse $7.0\times2.8$\,pc diam.; pos. angle $-30^{\circ}$; centred on RA 18:00:55.9 Dec -24:09:14.1 shown in Figure~\ref{fig:whole_velopix}}\\[-1mm]
\multicolumn{9}{l}{\scriptsize $^{e}$ For ellipse $5.6\times2.8$\,pc diam.; pos. angle $-10^{\circ}$; centred on RA 18:01:45.2 Dec -23:59:51.2 shown in Figure~\ref{fig:whole_velopix}}\\[-1mm]
\multicolumn{9}{l}{\scriptsize $^{f}$ For ellipse $2.8\times1.4$\,pc diam.; pos. angle $+30^{\circ}$; centred on RA 18:02:00.7 Dec -23:53:01.4 shown in Figure~\ref{fig:whole_velopix}}\\
\hline
\end{tabular}
\end{table*}


\section{Analysis and Results Overview}
\label{sec:overview}

Of the seventeen molecular transitions searched for
(Table~\ref{tab:lines}), eleven were detected. From the mapping data
we detected: CS, \C34S and \13CS in the (1-0) transition; SiO\,(1-0)
in several vibrational modes; CH$_{3}$OH; HC$_{3}$N\,(5-4) and
HC$_{5}$N in the (17-16) and (16-15) transitions.

\subsection{CS Emission}

The CS\,(1-0) line was detected across the W28 field and is the most
prominent. Along with the common CS isotopologue, $^{12}$C$^{32}$S we
also detect other isotopologues, \C34S and \13CS in the (1-0)
transition. Unless otherwise indicated, CS refers to the common
$^{12}$C$^{32}$S isotopologue. From our deep pointing observations
(DP-1 to DP-5, locations shown in Figure~\ref{fig:exposure}) with
\trms $\sim$2 times lower than the mapping data, a Gaussian fit to the
detected isotopologue line emission allowed the optical depth and
column density to be estimated. The method assumed an elemental
abundance ratio between CS and the rarer isotopologues of 22.5 for
[CS]/[\C34S] and 75 for [CS]/[\13CS] to calculate the optical depth
using~\citet[Equation 1]{zinchenko}. The upper state column density
was then estimated via the method from~\citet[Equation
  9]{goldsmith_langer}. The total column density of all states was
obtained by applying a local thermal equilibrium (LTE) approximation
and using temperature estimations from our \nh3
observations~\citep{me}, assuming that the CS gas is tracing the same
gas as our \nh3 observations. From these parameters, the LTE hydrogen
mass, number density and virial mass were estimated via the methods
outlined in~\citet{me}, by assuming a compact-source size $r=0.2$\,pc
and a molecular abundance ratio, {\Large $\chi$}$_{\rm CS}$, of CS to
H$_{2}$. CS molecular abundance ratios are known to vary dramatically
from different sites in the Galaxy, ranging from
$1\times10^{-8}$~\citep[eg. Sgr B - Table 4]{frerking} to
$6\times10^{-11}$~\citep[eg. W43 - Table 3]{linke}. We adopted an
abundance ratio of {\Large $\chi$}$_{\rm
  CS}=4\times10^{-9}$~\citep{irvine} following~\citet{zinchenko} which is 
typical for dense quiescent gas. This is likely the case for most of the clumps,
with the exception of the NE cloud as later discussed in section~\ref{sec:NEcloud}.
The results of our compact-source size LTE analysis and cloud
mass/densities from deep pointing observations are displayed in
Table~\ref{tab:massdens}.

The LTE analysis method has also been applied to larger extended
regions by utilising spectra from mapping data. For extended source
calculations, we extracted and averaged the spectra for all pixels
contained within an elliptical region. In this way, we apply the
standard analysis outlined above to a single averaged spectra for the
entire region. We used the same abundance ratio, but assumed the
emission fills a volume defined by a prolate ellipsoid of volume
$V=4/3 \times \pi\,r_{1}\,r_{2}^{2}$. Results of the extended source
LTE analysis from mapping data are displayed in
Table~\ref{tab:extdmassdens}

\subsection{SiO Emission}

We also tuned MOPS to include several different SiO\,(1-0) transitions
arising from different vibrational modes ($v=0,1,2,3$), as well as the
isotopologues $^{29}$SiO and $^{30}$SiO. The SiO molecule is a known
and proven signpost for the presence of shocked and disrupted
gas~\citep{Schilke:1996,Martin-Pintado:2000,Gusdorf:2008a,Gusdorf:2008b}.
Our detection of SiO from several locations provides some insight into
the disruptive conditions within the molecular clouds. Unless
otherwise indicated, SiO refers to $^{28}$SiO in the $v=0$ mode.

\subsection{CH$_{3}$OH Emission}

Multiple detections of CH$_{3}$OH masers towards the southern clouds
are indicative of the star formation activity present towards the \hii
regions and surrounding environments. Additional detections of
CH$_{3}$OH emission in the NE cloud is interesting. To date, no star
formation tracers such as: H$_{2}$O masers, CH$_{3}$OH masers or
infra-red emission have been detected towards this cloud. The
CH$_{3}$OH emission in the NE cloud does not appear to correspond to
stellar activity, as discussed later.

\subsection{Cyanopolyyne HC$_{n}$N ($n=3,5,7)$ Emission}

Detections of the cyanopolyynes HC$_{3}$N\,(5-4), HC$_{5}$N\,(16-15)
and HC$_{5}$N\,(17-16) towards the southern clouds are indicative
of {\bf activity in cold dense cores at the very early stages of high mass star 
formation}. Interestingly, additional detections of HC$_{3}$N\,(5-4) 
towards the shocked NE cloud are revealed. The sites of 
HC$_{3}$N\,(5-4) emission in the NE cloud do not appear to
correspond to stellar activity.

A more detailed discussion of the morphology and implications of the
molecular transitions detected from each of the mapped regions follows
in Section~\ref{sec:specdisc}

\section{Specific Discussion of Regions}
\label{sec:specdisc}

\subsection{North-eastern Cloud / HESS J1801-233}
\label{ssec:northeast}

Located at the northern boundary of HESS J1801-233 and nearby the
giant \hii region M20, we observe several molecular lines, which
reveal dense gas and star formation activity. These molecular lines
are all at a \vlsr range $\sim20$\,\kms, which is further away than
the rest of the molecular emission seen in the NE cloud, \vlsr
$\sim10$\,\kms, discussed later. This core was detected in our 12\,mm
study and where it was referred to as Core\,1. Toward this northern
most core is an IRAS source, IRAS 17589-2312~\citep{bronfman}, and we
adopt the IRAS nomenclature. The dense CS core seen here is likely
part of a molecular bridge connecting down towards the peak of the
HESS J1801-233 TeV emission seen in Figure~\ref{fig:exposure}. We note
that in the peak pixel map, Figure~\ref{fig:exposure}, the lowest
contour level is $\sim10$\,\trms. This gives an indication to the
prominence and the extent of the CS\,(1-0) emission detected towards
the NE cloud and the W28 clouds in general.  Toward IRAS 17589-2312,
we observe multiple Class I CH$_{3}$OH masers along the line of sight
within the beam. The maser emission is relatively weak (peak
\tmb$\sim$0.6\,K) compared to the other CH$_{3}$OH masers seen in the
W28 field discussed later. This detection adds to the list of maser
transitions which have been observed from this region: 6.7\,GHz
CH$_{3}$OH~\citep{fontani}, 22\,GHz H$_{2}$O~\citep{codella,me} and
now 44\,GHz CH$_{3}$OH.

We also detect the cyanopolyyne HC$_{3}$N\,(5-4) line towards this
core indicating hot gas phase chemistry probably driven by the central
source. There is also a detection of SiO\,(1-0) which indicates shocks
and/or outflows originating from IRAS 17589-2312. \citet{lefloch}
label this core as TC5 and suggest that the star formation in
this core is recent and could possibly have been triggered by the
W28 SNR. Spectra for the line emission detected towards this core
  are shown in Figure~\ref{fig:core1spec}.

Further south an extended dense region is spatially consistent with
the TeV gamma-ray source HESS J1801-233. Towards the north-east rim of
W28 several mm-line
surveys~\citep{wootten,arikawa,reach,torres,hess_w28,nanten21,me} have
revealed the well-known SNR shock interaction site, traced by broad
line measurements and the presence of many 1720\,MHz OH
masers~\citep{frail,claussen}.

\begin{figure*}
\includegraphics[width=\textwidth]{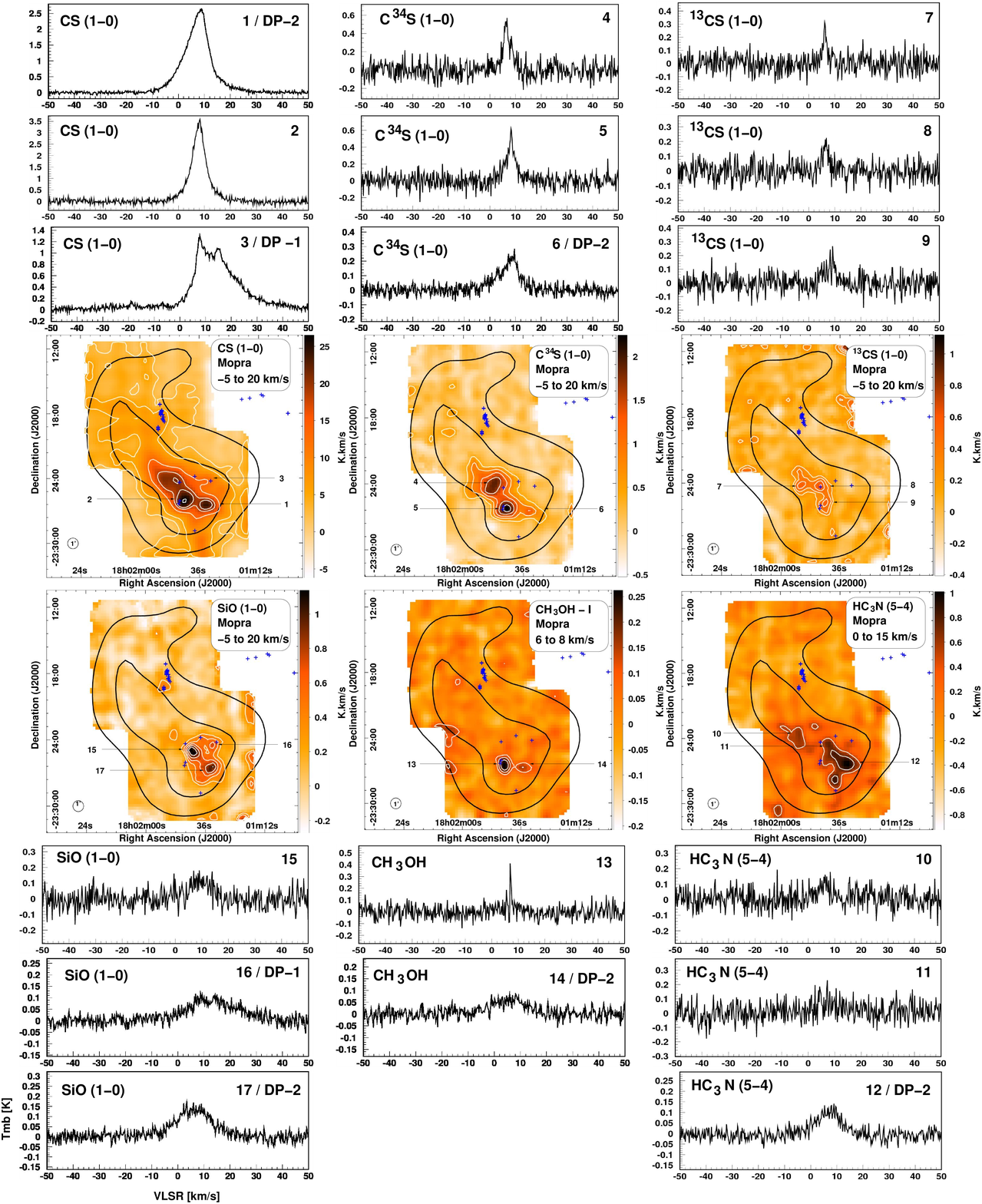}
\caption{Integrated intensity images and spectra for the north-eastern
  HESS J1801-233 region. Integration ranges are indicated on each
  panel. In all panels, H.E.S.S. TeV emission is indicated by thick
  black contours (4 \& 5$\sigma$ levels), the positions of 1720\,MHz
  OH masers from \citet{claussen} are indicated by blue crosses $+$,
  and white contours are used to highlight the molecular emission from
  the image. Minimum contour level for: CS\,(1-0) is 2.6\,\Kkms
  ($\sim$10\,\trms), for \C34S\,(1-0) is 0.4\,\Kkms ($\sim$2\,\trms),
  for \13CS\,(1-0) is 0.24\,\Kkms ($\sim$2\,\trms), for SiO\,(1-0) is
  0.32\,\Kkms ($\sim$2\,\trms), for CH$_{3}$OH 0.1\,\Kkms
  ($\sim$2\,\trms) and for HC$_{3}$N\,(5-4) is 0.5\,\Kkms
  ($\sim$2\,\trms).  Surrounding the integrated intensity images are
  spectra taken from the both the mapping data and deep pointings,
  labelled according to their locations.}
\label{fig:intmultipanel}
\end{figure*}

Figure~\ref{fig:intmultipanel} contains integrated intensity images
for emission detected towards this region.  The CS\,(1-0) emission is
spatially matched to the CO\,(2-1) emission measured by the Nanten
telescope~\citep{nanten21}, and \nh3 emission measured by Mopra
revealed a dense clump in the NE cloud~\citep{me}. This spatial
overlap is generally expected as CS will trace the denser
($n\sim$10$^{5}$\,\cmcube) regions of the molecular clouds. An image
showing the spatial relationship between the CO\,(2-1), CS\,(1-0) and
\nh3(3,3) gas is the NE cloud is shown in
Figure~\ref{fig:densediffuse}.  

The arcmin beam size allows the dense portion of the cloud to be
probed in more detail than in our two arcmin beam \nh3
study. Figure~\ref{fig:intmultipanel} highlights the clumpy nature of
the cloud. Several dense clumps are arranged in a north-east to
south-west direction. We also note the minimum contour level for the
integrated CS\,(1-0) emission in Figure~\ref{fig:intmultipanel} is set
to 10\,\trms. The CS\,(1-0) emission is very prominent throughout the
NE cloud and likely extends towards the limits of our mapping.

The line profiles of the CS and isotopologue emission are broad, with
FWHM\,$\sim$18\,\kms consistent with other molecular line observations
from the region. Additionally, the asymmetry in the spectra provide
evidence for the presence of shocks and gas
disruption~\citep{seta_1998}. Some CS line profiles indicate wings on
the `positive', or far side of the peak emission (e.g. CS\,(1-0) point
3/DP-1 on Figure~\ref{fig:intmultipanel}) and also on the `negative',
or nearer, side of the peak emission in others (e.g. CS\,(1-0) point
1/DP-2 on Figure~\ref{fig:intmultipanel}). In addition to the
asymmetric line profiles, symmetric line profiles (or at least
profiles with similar wings on both sides of the peak emission) are
also seen towards this NE cloud (e.g. CS\,(1-0) point 2 on
Figure~\ref{fig:intmultipanel}). These line profiles may suggest there
are components of the shock travelling in different directions, the
possibility of reverse shocks or may simply be a projection effect.

Several clumps of SiO\,(1-0) are detected and bounded by an arc of
1720\,MHz OH masers, further confirming that a shock has disrupted
this molecular cloud.  The peak of the SiO\,(1-0) emission (point 15
in Figure~\ref{fig:intmultipanel}) lies offset from the peak of the
dense gas traced by CS\,(1-0) (point 2 on
Figure~\ref{fig:intmultipanel}), which may suggest that the densest
CS\,(1-0) core is shielded from the disruption of the shock.

A class I CH$_{3}$OH maser transition is detected toward the eastern
edge of the dense gas (point 13 on
Figure~\ref{fig:intmultipanel}). This is the weakest of the CH$_{3}$OH
masers detected, with a peak temperature of $\sim0.4$\,K. A second
possible site of emission appears real from the DP-2 spectra (point
14/DP-2 on Figure~\ref{fig:intmultipanel}). Here the spectra is weak,
peak \tmb 0.06\,K, but with broad FWHM 15\,\kms. Both these CH$_{3}$OH
emission regions are not associated with typical star forming
cores. Rather, they appear to be a by-product of the SNR shock heating
the gas in a similar manner as discussed, for the case of expanding shocks from \hii
regions in~\citet{voronkov}.  Similarly, the detection of
HC$_{3}$N\,(5-4) from this region further indicates hot core gas phase
chemistry.  The emission is again broad, FWHM\,$\sim$12\,\kms as for
other lines detected from the region.
  
\begin{figure*}
\includegraphics[width=0.49\textwidth]{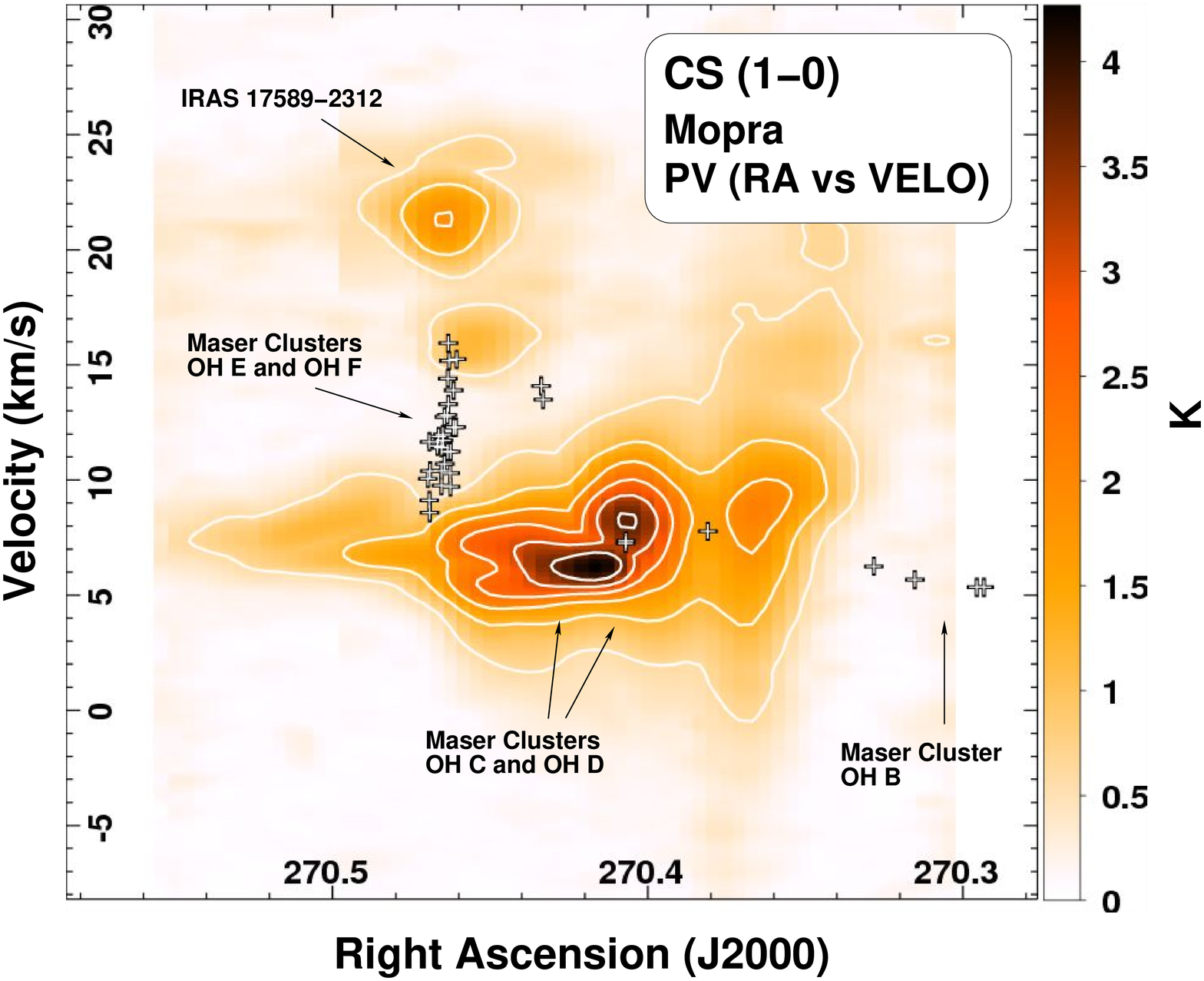}
\includegraphics[width=0.49\textwidth]{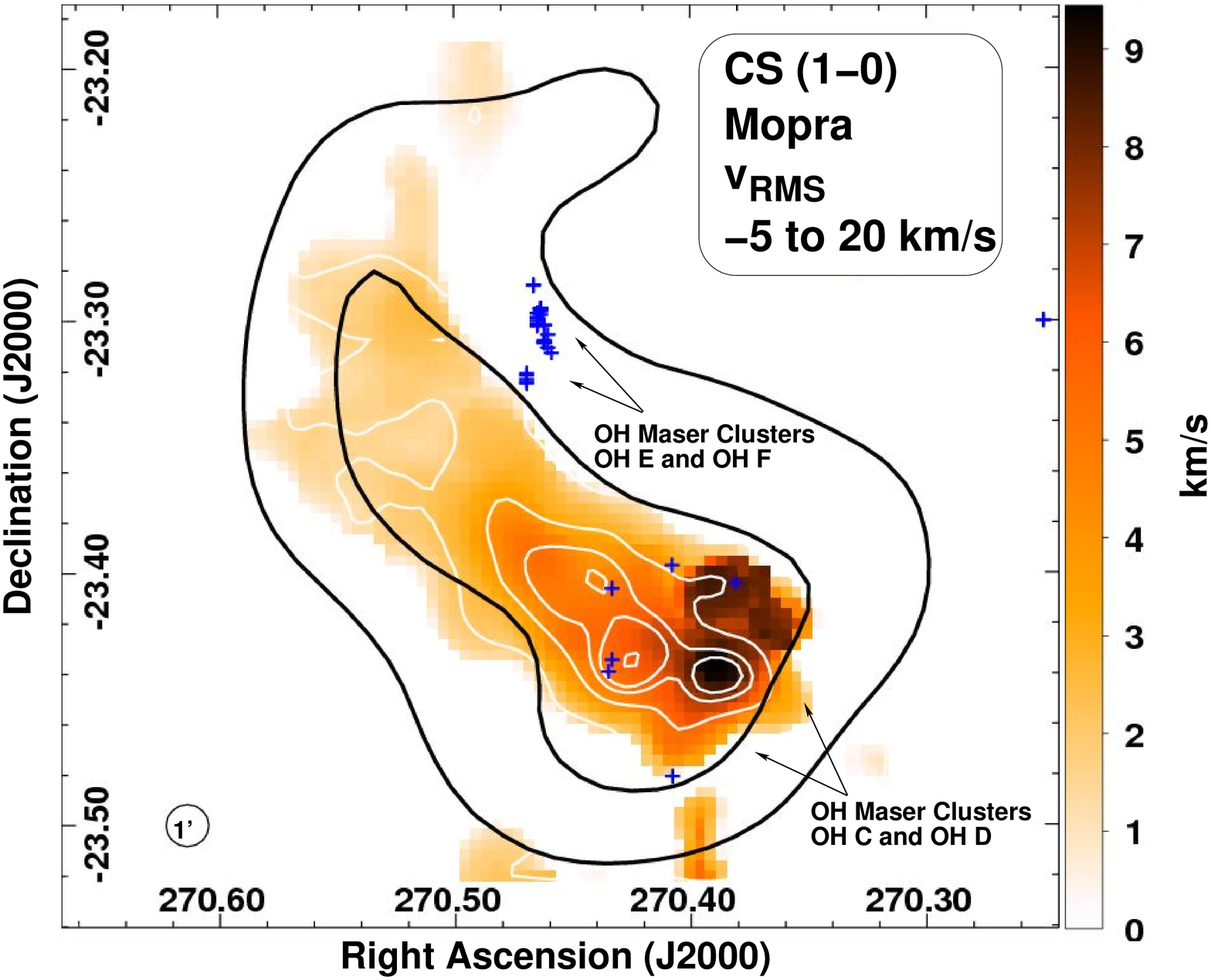}
\caption{{\bf Left:} Position velocity (PV) plot for the north-eastern
  region CS\,(1-0) emission. This RA vs \vlsr image is showing the
  peak pixel along the declination axis. Black/white crosses $+$
  represent the positions of the 1720\,MHz OH masers with their
  corresponding labels from~\citet{claussen}. The separate core at
  \vlsr $\sim20$\,\kms corresponds to the IRAS source. {\bf Right:}
  Velocity dispersion \vrms map for the CS\,(1-0) emission from the
  north-eastern region. The \vrms has been calculated for emission
  $\geq0.5$\,K in the -5 to 20\,\kms interval and then scaled to show
  the line FWHM. White contours are the integrated CS contours seen in
  Figure~\ref{fig:intmultipanel}. In both panels the RA axis is shown
  in degrees for clarity.}
\label{fig:core2pv}
\end{figure*}

The position velocity (PV) plot shown in Figure~\ref{fig:core2pv}
highlights the CS\,(1-0) emission, especially from the western or W28
side of the dense cloud. The velocity at the peak pixel map shown in
Figure~\ref{fig:whole_velopix} reveals the existence of a velocity
gradient, where the peak emission towards the south-western edge of
the cloud has larger \vlsr values which decrease towards the
north-east. This gradient would suggest that the shock is disrupting
the cloud from the south-western or W28 side.  The velocity dispersion
\vrms image for the CS\,(1-0) emission (for \tmb$>$0.5\,K in the -5 to
20\,\kms range) from the north-eastern region is also shown in
Figure~\ref{fig:core2pv}. The emission with the largest \vrms
dispersion originates from the south-western side of the cloud, which
was also the case for \nh3(3,3) emission in~\citet{me}. The more
disrupted emission is generally offset from the peaks seen in the
integrated map, which are also shown on the PV plot
(Figure~\ref{fig:core2pv}) as white contours. Interestingly one of the
clumps seen in the CS\,(1-0) integrated map (Point 1/DP-2 in
Figure~\ref{fig:intmultipanel}) appears disrupted and has a large
\vrms value. This clump also corresponds to a peak in the SiO\,(1-0)
integrated emission (Point 17/DP-2 in Figure~\ref{fig:intmultipanel}).

We have taken two position switched deep pointings towards the NE
cloud, specifically targeted towards the western side of the cloud
where the most turbulent conditions are seen. The locations of the
position switched observations are illustrated in
Figure~\ref{fig:exposure}. DP-1 is towards the broadest and most
disruptive region, and DP-2 is towards a disrupted core. From these
deep spectra, we have been able to extract gas parameters, which are
displayed in Table~\ref{tab:massdens}. We find that the compact-source
size mass, assuming a core with radius $r=0.2$\,pc, is more than a factor 10
smaller than the mass obtained when using the virial theorem. This is
to be expected as these observations are targeted to the most
disrupted regions of the cloud which are not expected to be
gravitationally bound. Additionally, the molecular abundance ratio may
be quite different in this shocked region compared to the unshocked gas.

\subsection{South-eastern Cloud / HESS J1800-240\,A}
\label{ssec:HESSJ1800-240A}

\begin{figure*}
  \includegraphics[width=\textwidth]{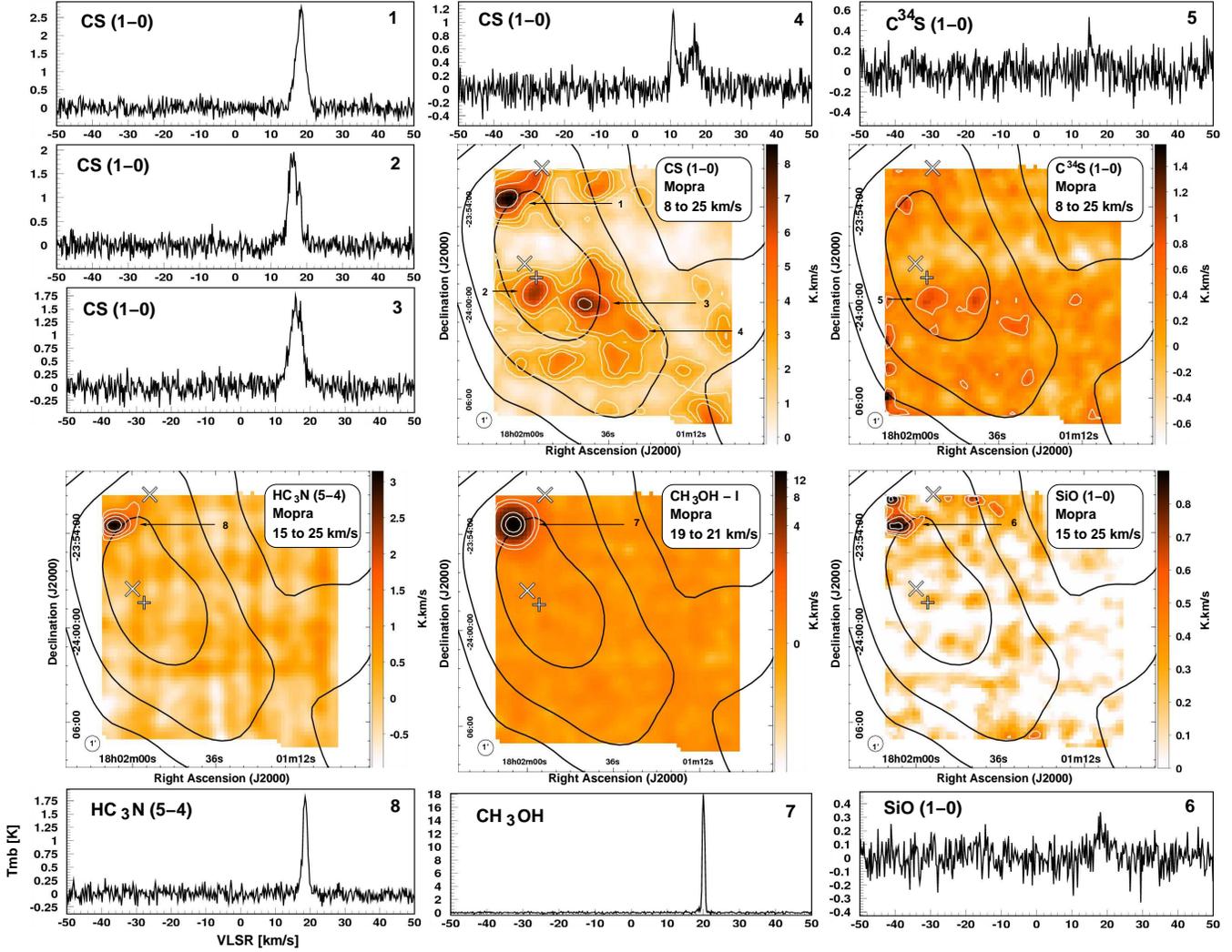}
  \caption{Integrated intensity images and spectra for the emission
    seen towards HESS J1800-240\,A. Integration ranges are indicated
    on each panel. In all panels, black contours are the H.E.S.S. TeV
    emission (4-6$\sigma$), black/white crosses $\times$ indicate the
    positions of the \hii regions G6.225-0.569 (north) and G6.1-0.6
    (south), the cross $+$ is the location of IRAS 17588-2358, and the
    white contours are used to highlight the molecular emission from
    the image. Minimum white contour level for CS\,(1-0) is
    1.75\,\Kkms ($\sim$5\,\trms); for \C34S is 0.52\,\Kkms
    ($\sim$2\,\trms); for HC$_{3}$N\,(5-4) is 1.04\,\Kkms
    ($\sim$4\,\trms); for CH$_{3}$OH is 0.22\,\Kkms ($\sim$3\,\trms)
    and for SiO\,(1-0) is 0.42\,\Kkms ($\sim$3\,\trms). Surrounding
    the integrated images are spectra taken from the mapping data and
    are labelled according to their locations.}
  \label{fig:core4int}
\end{figure*}

The region associated with HESS J1800-240\,A contains two \hii
regions, G6.225-0.569 and G6.1-0.6~\citep{lockman,kuchar} and IRAS
17588-2358. Integrated intensity maps of detected emission are shown
in Figure~\ref{fig:core4int}. Dense gas traced by CS\,(1-0) populates
this region in clumpy cores near the two \hii regions. A core
unrelated to the \hii regions (point 3 in Figure~\ref{fig:core4int})
shows signs of extension and also displays a double peaked line
profile towards its edge (point 4 in
Figure~\ref{fig:core4int}). Further south, additional clumps are
detected with $>5$ integrated \trms. The line profiles of the CS
emission towards these cores are also moderately broad, FWHM
$\sim5$\,\kms, but not to the extent of the emission seen in the NE
cloud.

In the NE corner of our mapping, near the \hii region 6.225-0.569, we
trace star formation activity, detecting a bright Class I CH$_{3}$OH
maser (peak \tmb$\sim$18\,K), a HC$_{3}$N\,(5-4) emitting core and
evidence of shocks/outflows with a detection of SiO\,(1-0). The
location of these star formation tracers are offset from the catalogued
position of G6.225-0.569.

We note that this region, associated with HESS J1800-240\,A, has less
exposure (a factor 3 lower than the north-eastern HESS J1801-233 and
southern HESS J1800-240\,B regions).

\subsection{Southern Cloud / HESS J1800-240\,B / G5.89-0.39}
\label{ssec:south}

\begin{figure*}
\includegraphics[height=0.92\textheight]{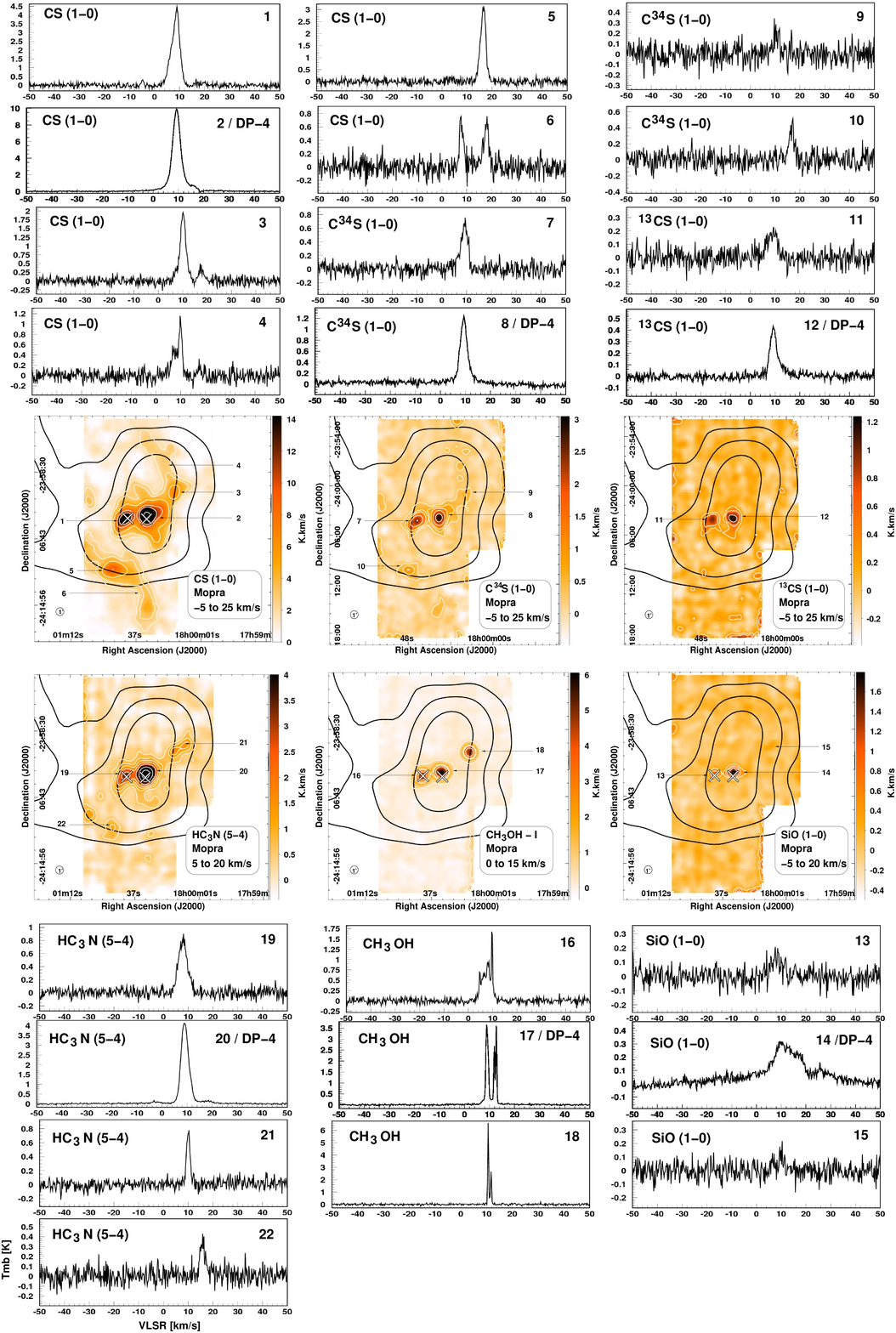}
\caption{Integrated intensity maps and spectra for the southern HESS
  J1800-240\,B region. In all panels, black/white crosses $\times$
  indicate the positions of G5.89-0.39\,A (left) \& B (right) and
  thick black contours are the H.E.S.S. TeV emission
  (4-6$\sigma$). White contours are used to highlight the molecular
  emission from the image. Minimum contour level for: CS\,(1-0) is
  2.08\,\Kkms ($\sim$8\,\trms); \C34S\,(1-0) is
  0.32\,\Kkms($\sim$2\,\trms); \13CS\,(1-0) is 0.24\,\Kkms
  ($\sim$2\,\trms); HC$_{3}$N\,(5-4) is 0.57\,\Kkms ($\sim$3\,\trms);
  CH$_{3}$OH-I is 0.5\,\Kkms ($\sim$5\,\trms) and SiO\,(1-0) is
  0.36\,\Kkms ($\sim$3\,\trms). Surrounding the integrated images are
  spectra taken from both the mapping data and deep pointings,
  labelled according to their locations.}
\label{fig:southCS}
\end{figure*}

Spatially well matched to HESS J1800-240\,B are three dense cores
oriented in a south-east (SE) to north-west (NW) alignment (points 1,
2 \& 3 in Figure~\ref{fig:southCS}), which were seen in our previous
12\,mm observations~\citep[labelled as Triple Core SE, Central and
  NW]{me}. The SE and central sources are associated with the
IR-bright and energetic UC \hii region G5.89-0.39 and the NW source
with a pulsating M type star, V5561 Sgr, a further 5 arcmin away.

The extraordinary molecular outflow emanating from the \hii complex
G5.89-0.39 has been extensively studied from arcsec to arcmin scales
in multiple molecular lines
(e.g.~\citet{harvey,churchwell,zijlstra,gomez,choi,acord,claussen,thompson,kimkoo2001,kimkoo2003,sollins,hunter,me})

Outflow velocities $\pm70$\,\kms are seen in CO lines and similar
claims are made for SiO lines by~\citet{acord}. \citet{sollins} argues
that the outflow originates from a 1.3\,mm continuum source and not
from Feldt's candidate star~\citep{feldt}, as the star is not
equidistant from the outflow lobes. The \hii complex G5.89-0.39 is
actually comprised of two active star formation regions, \hii
G5.89-0.39\,A to the east, and UC\hii G5.89-0.39\,B approximately 2
arcmin to the west~\citep{kimkoo2001}. G5.89-0.39\,A is also known as
W28\,A2 with strong radio continuum emission, while G5.89-0.39\,B
possesses the outflow.  We observe SiO\,(1-0) over a wide region
towards the three sources suggesting there are shocks or
outflows. Multiple higher order SiO transitions on smaller scales are
detected towards the G5.89-0.39\,B region and have been discussed
by~\citet{acord}.

A new perspective that this work provides is the broad scale (15
arcmin) dense gas surrounding the \hii regions. NW of the UC \hii
region G5.89-0.39\,B in CS\,(1-0) we see an extended arm feature with
length $\sim6$ arcmin extending north towards W28 (e.g points 3 \& 4
in Figure~\ref{fig:southCS}). The dense gas seen in this arm appears
to have a slight velocity gradient, approximately
$+0.3$\,\kms\,arcmin$^{-1}$ from north to south. To the SE of the \hii
regions there are several dense clumps which trace the dense component
of the cloud seen in CO\,(2-1) by~\citet{nanten21}. Once again, these
cores display broad CS\,(1-0) emission (FWHM$>$4\,\kms) however, these
clumps are at a slightly larger \vlsr compared to the \hii
regions. Interestingly, most of the CS\,(1-0) emission from the
southern cloud contains multiple clumps along the line of sight, all
within $\sim$20\,\kms of each other. This makes \vrms calculations
difficult as the multiply clumped emission produces artificially broad
results.

Star formation tracers such as 44\,GHz class I CH$_{3}$OH masers and
HC$_{3}$N\,(5-4) emission have been detected towards all three of the
aligned sources. We also detect signs of warm gas phase chemistry via
detections of HC$_{3}$N\,(5-4) towards dense cores SE of the \hii
regions (e.g point 22 in Figure~\ref{fig:southCS}) which lie in a
dense molecular ridge (see Figure~\ref{fig:exposure}). We also detect
the HC$_{5}$N\,(16-15) and (17-16) lines towards G5.89-0.39\,B. This
would indicate that there is considerable energy available to excite
the molecules into such a highly excited state.

An additional site of stellar activity is seen towards our
southernmost mapped region. At the base of the dense ridge seen in
CS\,(1-0), we detect SiO emission in the $v=1,2,3$ modes of the (1-0)
line. Here we are probably tracing the radiationally excited SiO
emission from the variable star V5357\,Sgr. Spectra for the SiO
emission are displayed in Figure~\ref{fig:SiO123}.

 We have also taken position switched deep pointings towards
 G5.89-0.39\,B (DP-4) and towards the northern most point of the
 arm NE of G5.89-0.39 (DP-3). We have estimates of the mass and density from
 the CS isotopologue emission towards G5.89-0.39\,B. We do not detect
 any isotopologue emission from the extended arm position (DP-3) so
 we estimate the gas parameters under the assumption that the
 CS\,(1-0) emission is optically thin. The results of the compact-source
 analysis are displayed in Table~\ref{tab:massdens}.

\subsection{HESS J1800-240\,C / G5.71-0.08}
\label{ssec:OHmaser}

\begin{figure*}
\includegraphics[width=0.95\textwidth]{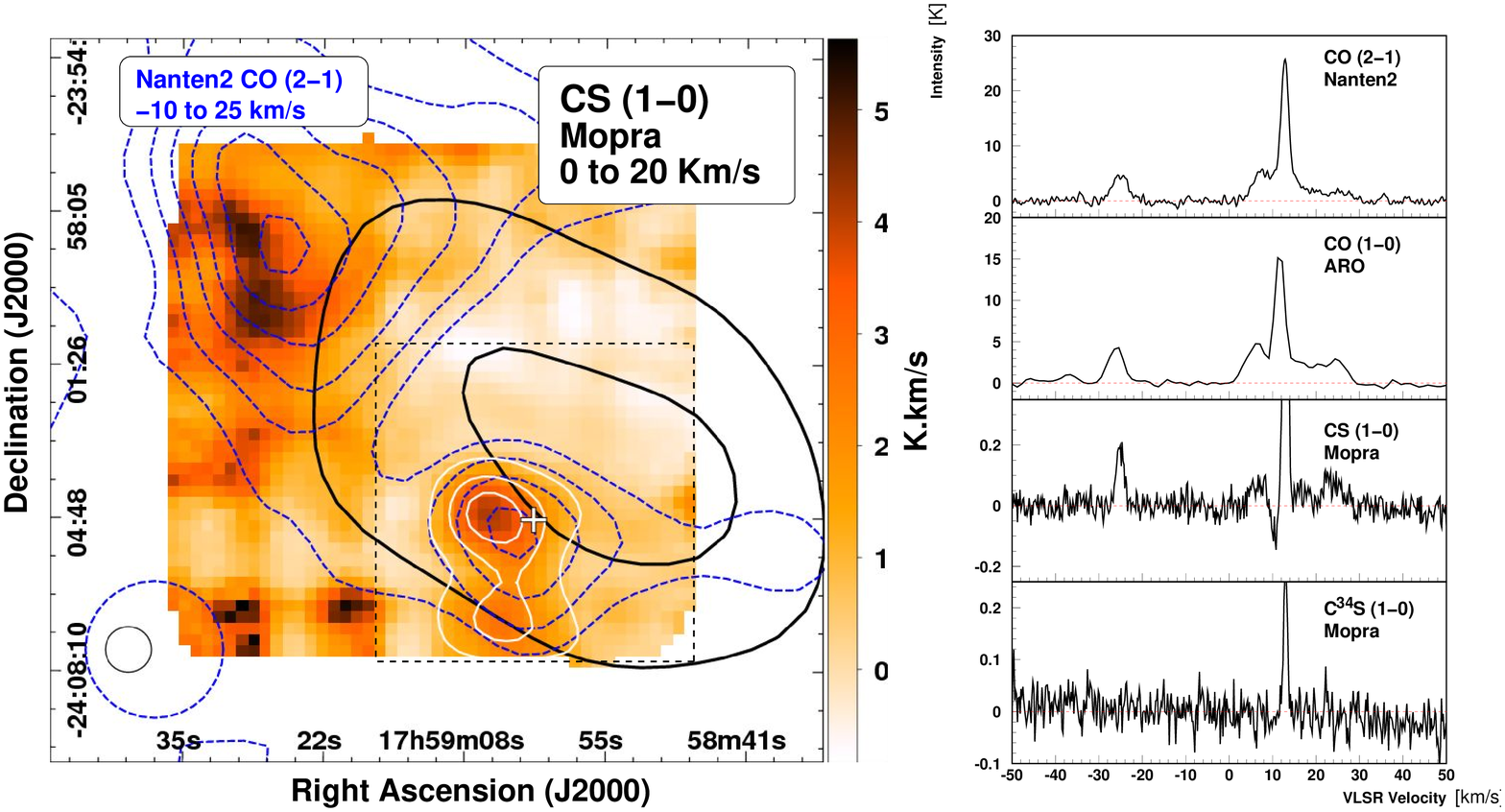}
\caption{{\bf Left:} Integrated intensity image of CS\,(1-0) emission
  towards HESS J1800-240\,C, which contains the recently detected
  1720\,MHz OH maser (G5.7-0.0) by~\citet{hewitt}. The maser position is
  indicated by the cross $+$. Blue dashed contours are Nanten CO\,(2-1)
  integrated intensity contours from~\citet{nanten21}. White
  contours are used to highlight CS\,(1-0) emission, only for the region of highest
  exposure (indicated by the black dashed box. Minimum white
  contour level is 0.88\,\Kkms ($\sim$2\,\trms). {\bf Right:} Spectra
  for the CO\,(2-1)~\citep{nanten21} CO\,(1-0)~\citep{liszt},
  CS\,(1-0) and \C34S\,(1-0) for the maser position. Spectra for the
  CS isotopologue emission are taken from the Mopra deep pointing at
  the location of the OH maser. For the CS\,(1-0) emission, the
  vertical scale has been clipped (peak temperature 2.5\,K) to show
  the evidence for line-wings and absorption in the profile.}
\label{fig:OHmaser_multi}
\end{figure*}

The CS\,(1-0) emission detected towards HESS J1800-240\,C shows two
emitting regions along the line of sight, one with
\vlsr$\sim-25$\,\kms and another with \vlsr$\sim13$\,\kms. This dense
CS core is slightly offset from the peak of the TeV source and may
provide a suitably dense region for CR interactions. The CS emission
is also spatially matched with the SNR candidate G5.71-0.08 and the
1720\,MHz OH maser at \vlsr$=13$\,\kms~\citep{hewitt}. The OH maser
may indicate that HESS J1800-240\,C is tracing another SNR/molecular
cloud interaction. Figure~\ref{fig:OHmaser_multi} is an integrated
intensity image of the HESS J1800-240\,C region comparing our
CS\,(1-0) emission and the CO\,(2-1) emission from~\citet{nanten21}.
Spectra from our \C34S\,(1-0) and CS\,(1-0) deep pointings, the
CO\,(2-1) spectra from~\citet{nanten21} and CO\,(1-0) spectra
from~\citet{liszt} at the position of the 1720\,MHz OH maser are also
presented.

The spectra presented in Figure~\ref{fig:OHmaser_multi} reveal
line-wings in the CO\,(2-1) and (1-0) profiles, with \vlsr $\sim$0-8
and 20-30\,\kms, indicative of turbulent conditions. The CS\,(1-0)
profile also shows some structure in these \vlsr ranges, which could
be interpreted as multiple clumps along the line of sight, or evidence
for line wings. There is also an absorption dip at \vlsr
$\sim10$\,\kms. Similarly, absorption features are present in the
1720\,MHz OH spectrum at \vlsr$=7$ and $-25$\,\kms~\citep{hewitt}, and
may also be evident in the CO emission. This is likely due to
absorption against a continuum source. Interestingly the width of the
CS\,(1-0) line profile is narrow, with FWHM 1\,\kms. This could
suggest the central region of the core is rather quiescent.

The two isotopologues of CS are detected in our deep position switched
observation towards the new OH maser position and allow the mass and
density to be estimated. For this CS core we estimate a mass of
190\,M$_{\odot}$ and a H$_{2}$ number density of
1.7$\times10^{5}$\,\cmcube.
Finally, we note that this mapped region associated with HESS
J1800-240\,C has less exposure (a factor 1.5 to 3 lower) than the HESS
J1801-233 and HESS J1800-240\,B regions.

\section{X-Ray \& TeV Gamma-Ray Emission and discussion}
\label{sec:HEdiscussion}

\subsection{HESS J1801-233/NE Shocked Cloud}
\label{sec:NEcloud}

The CS\,(1-0) emission seen towards the north-eastern cloud traces an
extended dense region of molecular material. Using the emission within
a region defined by $\int~\tmb~dv~\geq~10$\,\Kkms ($\sim40$\,\trms)
and finding the average CS\,(1-0), \C34S\,(1-0) and \13CS\,(1-0)
emission profiles within the region, we can estimate the gas
parameters as an extended source. We apply this strict limit on the
CS\,(1-0) emission to only accept pixels containing CS, \C34S and
\13CS emission, and not bias the optical depth result. Under these
conditions, we estimate the CS emission fills an ellipse on the sky
with physical dimensions of $3.1\times1.4$\,pc (at the estimated
2\,kpc distance). This elliptical region is indicated on
Figure~\ref{fig:whole_velopix}. Additionally, we assume the volume of
the emission is defined by a prolate ellipsoid, where the third axis
(in the $z$ direction) is the same as the minor axis of the ellipse on
the sky. In this way, the ellipsoid has radii
$3.1\times1.4\times1.4$\,pc. Using the method outlined in
Section~\ref{sec:overview}, we calculate the mass
using~\citet[Equation 3]{me} and include an additional factor
$\eta_{\rm mb}/\eta_{{\rm xb}}=0.77$ to account for the Mopra extended
beam efficiency $\eta_{\rm{xb}}=0.43$~\citep{mopra_beam}.  This gives
an estimated extended source mass of $5.6\times10^{4}$\,\msun and a
H$_{2}$ number density $6.1\times10^{4}$\,\cmcube. The extended source
mass and density estimates are presented in
Table~\ref{tab:extdmassdens}.

We compare our CS gas mass to other mass estimates of the region. Over
the 0-12\,\kms velocity range, the CO\,(1-0) mass has been estimated
to be $\sim2\times10^{4}$\,\msun, and over the wider 0-25\,\kms
velocity range $\sim5\times10^{4}$\,\msun~\citep{hess_w28}. From
CO\,(3-2) data~\citet{arikawa} estimated there to be
$2\times10^{3}$\,\msun of shocked gas in the NE cloud and
$4\times10^{3}$\,\msun of unshocked gas. Additionally a conservative
lower limit to the mass of gas traced by \nh3 inversion transitions
has been estimated to be $>1300$\,\msun~\citep{me}.

At a glance the mass derived from CS is larger than previously published
estimates. It is at least 10$\times$ the lower limit obtained via \nh3
observations.  This is likely due to the considerable uncertainties in
molecular abundance ratios, which are usually the largest sources of
systematic error in mass/density calculations. While an order of
magnitude mass discrepancy is not ideal, it has been shown for a
sample of molecular cores, both hot and cold, that the core masses
have equal or higher values from CS observations than those found via
\nh3; in some cases discrepancies up to two orders of
magnitude~\citep{zhou}. We suggest the CS mass be an upper limit as
the abundance in a shocked source is likely to be lower than that we
have assumed applies in an undisturbed, dense molecular cloud. 
The fact that the morphological extent of the CS emission is similar to that in CO
however, suggests that a quite high fraction of the molecular gas is in a high
density state.

The X-ray morphology of W28 is described as centre
filled~\citep{rho2002} whereby X-ray emission lies towards the centre
of the SNR. Interestingly, there is a brightening of the X-ray
emission towards the NE of the SNR. This so called `{\it ear}'
feature, labelled by~\citet{rho2002}, is found at the boundary of the
pre- and post-shocked molecular cloud seen in CO\,(1-0) and
(3-2)~\citep{arikawa}. 

XMM-Newton X-ray observations towards the NE shocked cloud have been
discussed in~\citet{ueno2003} following earlier observations with ASCA
and ROSAT~\citep{rho2002}. Here, we have looked at additional
XMM-Newton archival observations totalling $\sim$56\,ks of observation
time. Data were reprocessed using the XMM SAS software suite (with
standard quality cuts) and major proton flares were filtered
out. Observation IDs used were 0135742401 (5.5\,ks post filter),
0145970101 (20\,ks - analysed by~\citet{ueno2003}), and 0145970401
(30\,ks).

\begin{figure*}
\centering
\includegraphics[width=\textwidth]{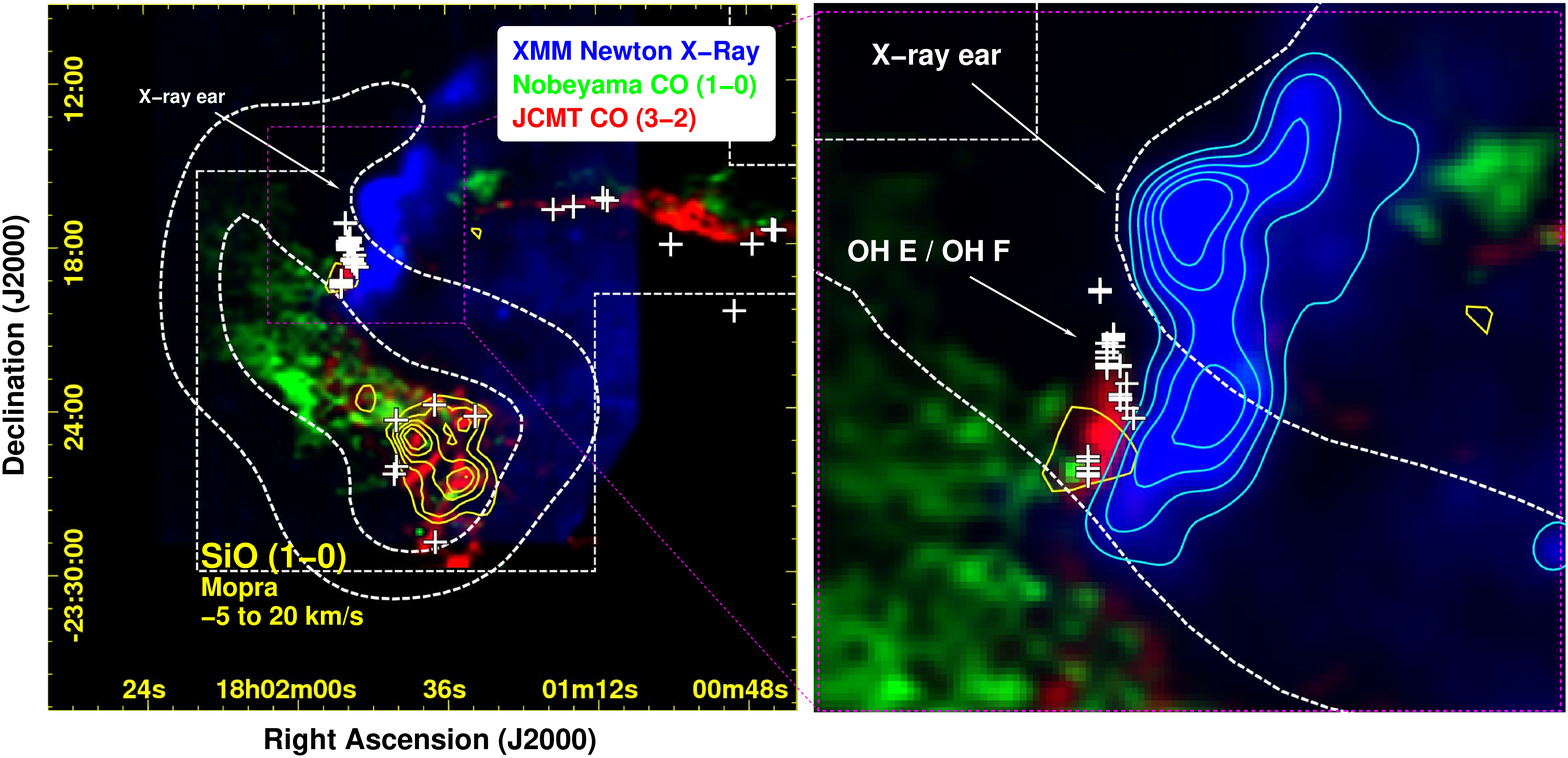}
\caption{{\bf Left:} Three colour image for the HESS J1801-233
  region. The red channel is JCMT CO\,(3-2) integrated from -40 to
  40\,\kms~\citep{arikawa}. The green channel is the Nobeyama
  CO\,(1-0) emission integrated from 4 to 9\,\kms~\citep{arikawa}. The
  blue channel is the XMM-Newton 0.2 to 10\,keV
  [ph\,\cmsqr s$^{-1}$] image (Gaussian smoothed). White
  dashed lines indicate the limits of the CO images. Thick white
  dashed contours are H.E.S.S. TeV emission (4 \& 5$\sigma$
  levels). Yellow solid contours are Mopra SiO\,(1-0) integrated from -5 to
  20\,\kms, as seen in Figure~\ref{fig:intmultipanel}. White crosses
  $+$ indicate the positions of 1720\,MHz OH masers
  from~\citet{claussen}. {\bf Right:} Zoom of the X-ray ear and OH
  maser clusters OH E and OH F from~\citet{claussen}.}
\label{fig:xray}
\end{figure*}

The X-rays result mostly from thermally heated gas (post shocked) with
$kT=0.3$\,keV~\citep{ueno2003}. Figure~\ref{fig:xray} presents the
XMM-Newton full EPIC camera (PN+MOS1+MOS2) image in the 0.2 to 10 keV
band corrected for exposure in the blue channel. The red and green
channels are the post-shocked CO\,(3-2) and pre-shocked CO\,(1-0) gas
from~\citet{arikawa}. The right panel of Figure~\ref{fig:xray} is a
zoom of the X-ray `{\it ear}' feature, revealing the boundary between
the X-ray emission and the shocked gas.  Also seen at this boundary
are many 1720\,MHz OH masers~\citep[groups OH E \& OH F, labelled on
  Figure~\ref{fig:core2pv}]{claussen}. Interestingly, adjacent to the OH maser groups E and F we also
detected SiO\,(1-0) at the 2\,\trms level, further enhancing the evidence that there is shocked
and disrupted dense gas in the region.

The cluster of masers (OH E \& OH F) are possibly associated with the
X-ray {\it `ear'} (Figure~\ref{fig:xray}). What is particularly
intriguing is that the OH E \& F clusters of 1720\,MHz OH masers
appear to lie on both sides of the shocked gas (Figure~\ref{fig:xray},
red channel) with respect to W28. This may result from projection effects which are hard
to disentangle or the possibility of an additional or reverse shock
propagating inward towards W28. The OH E \& F maser clusters lie in the 10-15\,\kms
velocity range (Figure~\ref{fig:core2pv}). This suggests that the
post-shocked X-ray emitting gas may lie behind most of the NE dense molecular cloud,
which peaks in the 5-10\,\kms velocity range (Figure~\ref{fig:core2pv}).

The broad molecular line emission observed towards the NE cloud would
likely result from non-thermal components. The additional kinetic
energy required to produce the broad line emission can be
calculated. We determine the non-thermal energy as $W_{\rm
  kin}~=~1/2M(\Delta v_{\rm kin})^{2}$, where $M$ is the mass of the
broad line gas, and $\Delta v_{\rm kin}$ is the line FWHM. Using the
average CS\,(1-0) line FWHM=7.3\,\kms and the average extended mass
upper limit $M=5.6\times10^{4}$\,\msun, we obtain $W_{\rm
  kin}<3.8\times10^{49}$\,erg, which is a few percent of the total
$10^{51}$\,erg released in a typical SNR explosion. This upper limit
is in agreement with our earlier estimate of the kinetic energy
deposited into the dense NE cloud ($W_{\rm kin}>7\times10^{47}$\,erg)
from our earlier \nh3 observations. Similarly,~\citet{arikawa} estimated
$\sim3\times10^{48}$\,erg of kinetic energy has been deposited into
the 2000\,\msun traced by shocked CO\,(3-2).

The impact of CRs and SNR shocks on molecular clouds has been
discussed extensively. \citet{aharonian1994} discuss a 
molecular cloud (MC) being overtaken by a SNR shock. The 
`crushed cloud' model from~\citet{uchiyama}
discusses the re-acceleration of particles within a molecular cloud, but
is mainly applicable to GeV energies. \citet{yamazaki2006} discuss
three cases for VHE emission from old or evolving SNRs and molecular clouds (MC). This includes (i)
emission from the SNR shock alone, (ii) the SNR shock colliding with a
MC and (iii) CRs accelerated a distance away from a MC and diffusing
through the interstellar medium (ISM) to illuminate the cloud. This
latter case of CR acceleration and diffusion is also more recently
discussed in~\citet{fermi_w28,gabici2010,ohira_2010}, with discussion
specifically toward W28.~\citet{fujita2009} discuss the scenario of
CRs accelerated by a SNR shock nearby or just outside the W28 MCs.
Recent work by~\citet{inoue} models the propagation of middle aged SNR
shocks into a cloudy ISM, revealing the
presence of multiple secondary shocks with total surface area
comparable to the primary shock. These 
shocks are able to accelerate particles to a maximum energy $E_{{\rm
    max}}>10$\,TeV~\citep{inoue}. Overall, all models can account 
for the TeV and GeV emission seen towards W28 and other similarly 
aged SNRs. They clearly favour the hadronic emission mechanism 
for TeV emission.

Under a hadronic emission mechanism, TeV
photons result from CR collisions producing $\pi^{0}$ mesons. The
$\pi^{0}$s then decay into the TeV photons. The leptonic emission
mechanism produces TeV photons by inverse Compton processes and/or
electron non-thermal Bremsstrahlung. However, these leptonic processes are
suppressed in evolved SNRs due to electron cooling via synchrotron losses
at much earlier epochs.

TeV gamma-rays are observed with energies
in the 0.3-3\,TeV range~\citep{hess_w28}. In the hadronic scenario, typically 17 percent of the CR
particle's energy is transferred to gamma-rays, therefore 0.3-3\,TeV
gamma-rays trace a parent population of $\sim2-20$ TeV CRs at the
source. The striking spatial overlap between the TeV emission and the
dense gas towards HESS 1801-233, suggests that CRs are reaching the
dense portions of the molecular cloud, and could help provide
constraints for CR diffusion models~\citep{gabici}.
However present H.E.S.S. observations do not have sufficient angular
resolution ($\sim10$\,arcmin FWHM) to test whether the CR spectrum 
varies towards the interior of the dense cores. 
Future TeV instruments with sufficient angular resolution (arcmin) are
required before such spectral features may become apparent and help to
discriminate between models of re-acceleration and diffusion. A future
TeV instrument, such as the Cherenkov Telescope Array
(CTA)~\citep{CTAdesign}, will have both improved angular resolution
and sensitivity. The expected angular resolution could reach a few arcmin FWHM 
and the
energy sensitivity is expected to be a factor of 10 better than
H.E.S.S., achieving $\sim$few $\times10^{-14}$\,erg\,\cmsqr\,s$^{-1}$
in $\sim50$\,hr. Together, these improvements will permit TeV emission
maps and detailed spectra with comparable spatial scales to the
molecular gas maps and provide a platform to test CR diffusion models
and probe re-acceleration models.

The expected flux of TeV emission from a molecular
cloud impacted by Galactic CRs (GCR) with particle flux following an integral
power law $E^{-1.6}$ can be found from~\citet[Equation 10]{aharonian_1991}:

\begin{equation}
F(\geq E_{\gamma})=2.85\times10^{-13}\,E^{-1.6}_{{\rm TeV}}\,\left( \frac{M_{5}}{d^{2}_{{\rm kpc}}} \right) \,k_{{\rm CR}} \hspace{2mm}{\small [{\rm ph\,cm}^{-2}\,{\rm s}^{-1}]}
\label{eqn:TeVFlux}
\end{equation}

\noindent where $M_{5}$ is the mass of the cloud (in $10^{5}$\,\msun),
$d_{{\rm kpc}}$ is the distance to the cloud (kpc) and $k_{{\rm CR}}$ is the
CR density enhancement factor above the local solar-system value.
The observed TeV spectral indices~\citep{hess_w28} imply
an incident proton particle spectral integral index of $\Gamma \sim 1.6$
(in a power law particle spectrum $\propto E^{-\Gamma}$) similar to
GCRs. Thus Equation~\ref{eqn:TeVFlux} should be sufficient for initial 
estimates of the TeV fluxes arising from parts of the clouds. We note that further
discussion of other systematic uncertainties concerning the
predicted TeV fluxes from cloud clumps is left for section~\ref{sec:conc}.
 
The CR enhancement
factors have been estimated towards three of the four TeV sources
in~\citep[Table 2]{hess_w28}. With this in mind, we can estimate the
gamma-ray flux for emission from the dense cloud components traced by
CS\,(1-0) emission. Using the NE shocked cloud extended source mass
estimate from Table~\ref{tab:extdmassdens}, we estimate
$F(\geq\,1\,{\rm TeV}) = 5.1\times10^{-13}$\,ph\,\cmsqr\,s$^{-1}$. The
flux results for this, and other CS regions, are summarised in 
Table~\ref{tab:highenergy}. We note
that this flux is $\sim15$\% higher than that 
currently detected by H.E.S.S. This further supports our
case for treating the CS mass estimate as an upper limit given the uncertainties in the CS abundance. 
The general picture though is that the predicted TeV flux is detectable and can be well
studied by CTA in a reasonable observation time (10-50\,hr). 
CTA could then discriminate between the TeV emission from within the
dense cloud component vs. that of the entire cloud as seen by
H.E.S.S..

\subsection{HESS 1800-240\,A \& B}

The southern TeV sources HESS J1800-204\,A and B are also associated
with dense gas which could provide a suitable target for CR
interactions. The dense regions of gas spatially consistent with HESS
1800-240\,A are treated as elliptical extended sources
filling prolate ellipsoids with radii $1.4\times 0.7\times 0.7$\,pc
for the gas nearby G6.225-0569 and $2.8\times 1.4\times 1.4$\,pc for
the gas nearby G6.1-0.6.
Similarly, the CS gas seen towards HESS J1800-240\,B is treated as
three separate extended sources. The large
clump associated with the G5.89-0.39 \hii regions is assumed
to fill a prolate ellipsoid with radii $2.4\times 2.1\times 2.1$\,pc.
The CS emission from the NW arm, extending from V5561 Sgr, is assumed
to fill an ellipsoid with radii $3.1\times 1.1\times 1.1$\,pc, and the
SE molecular ridge is assumed to fill an ellipsoid with radii
$3.5\times 1.4\times 1.4$\,pc. These elliptical regions are indicated
on Figure~\ref{fig:whole_velopix}. We detect only CS\,(1-0) emission
from the NW arm and SE ridge regions, therefore, we assume the
emission to be optically thin. The individual estimates for the mass
and density of all the extended regions towards HESS J1800-240\,A \& B
are summarised in Table~\ref{tab:extdmassdens}.

The three extended regions towards HESS J1800-240\,B suggest that
$7.1\times10^{4}$\,\msun of material is contained in dense clumps.
The HESS J1800-240\,B southern cloud is estimated to have
$10^{5}$\,\msun based on the CO\,(1-0) emission~\citep{hess_w28}. 
Based on our extended CS clump estimates,
we find that $\sim70$\% of the southern clouds mass is contained in
the dense clumps surrounding the G5.89-0.39 \hii regions. 

As these three dense extended regions all lie within the HESS
J1800-240\,B source, the improved angular resolution of CTA could
potentially resolve this emission into three separate sources. In this
case, the expected gamma-ray flux can be found via
Equation~\ref{eqn:TeVFlux}. The expected TeV flux for photon energy $E
> 1$\,TeV for each of the dense cloud components are summarised in
Table~\ref{tab:highenergy}. Typically the expected flux is
$10^{-13}$ to $10^{-14}$ ph\,\cmsqr s$^{-1}$, which could be detectable
and well studied by CTA in a reasonable observation time (10-50\,hr).
As for the NE cloud, we find that the total expected flux from the three HESS J1800-240\,B
sources is higher than the detected level of TeV emission (by a factor of $\sim$2). Again,
this could be explained by our choice of CS abundance being overestimated in this region which may be
expected given likely internally shocked nature of the G5.89-0.39 \hii complex. Additionally, our 
cloud-averaged CR density $k_{\rm CR}$ may not reflect smaller scale variations. 

The question of the origin of the CR population towards HESS J1800-240\,A and B is still
uncertain, although diffusion of CRs from W28 seems to reasonably well explain the observed
GeV to TeV gamma-ray spectra (e.g. \citet{fujita2009,fermi_w28,gabici2010,li2010}). However,
additional CRs accelerated inside the local star formation regions (from G5.89-0.39 in particular), and another potential
SNR towards HESS J1800-240\,C (discussed shortly) remains a possibility (see e.g. \citet{araudo}). 
Our observation of the arcmin-scale dense NW arm and SE ridge features in the molecular cloud may signal additional
sites of disruption which could arise from internal (star formation) and/or external (e.g. SNR shocks) forces. 
Further tests of these scenarios can come from future arcmin-scale studies of gamma-ray spectra 
throughout the clouds.

\subsection{HESS J1800-240\,C}

Towards HESS J1800-240\,C, we have discovered a dense molecular core.
Unlike the other TeV sources in the W28 region, the dense gas lies
slightly offset from the peak of the TeV emission.  The CS gas
is found in a dense, compact core, which has a narrow line width, FWHM
1\,\kms. Additionally, there is absorption in the CS line profile at
$\sim$10\,\kms, which may also be evident in the CO\,(1-0) and (2-1)
line profiles (see spectra in Figure~\ref{fig:OHmaser_multi}). 

Based on the CO\,(1-0) emission from the NANTEN
survey~\citep{nanten10}, we estimated the mass of the CO cloud. We
assumed a CO abundance of {\Large $\chi$}$_{\rm CO}$ =
1.5$\times10^{20}$ and assumed a spherical emission region with radius
0.1$^{\circ}$ centred on the HESS J1800-204\,C TeV source. We
included emission from the 0-20\,\kms range, which was selected to be
consistent with the ranges used in~\citet{hess_w28}. Under these
assumptions, we estimated 1.4$\times10^{4}$\,\msun of CO gas towards
HESS J1800-240\,C. We estimated the CR enhancement factor $k_{\rm CR}
= 35$, using Equation~\ref{eqn:TeVFlux}, based on the CO mass and the
detected TeV flux following~\citet{hess_w28}. Assuming the value of
$k_{\rm CR}=35$ we determined the expected TeV flux from just the
dense CS core. The result is shown in Table~\ref{tab:highenergy}

The determined CR enhancement of 35 is higher than towards the other
TeV sources.  Enhancements of this level are expected in the regions
of CR accelerators, which could be expected as this region could be
associated with both the W28 SNR as well as the local SNR candidate
G5.71-0.08 which may be producing the OH maser. Under a purely 
hadronic emission scenario, the expected
TeV flux from just the dense CS core is $F(E \geq 1\,{\rm
  TeV})\sim2\times10^{-15}$ ph\,\cmsqr s$^{-1}$.  This flux level
would likely push the reasonable detection limits for CTA.

\begin{table}
\caption{Summary of the CR enhancement factors, $k_{{\rm CR}}$, and
  the predicted TeV fluxes from the extended dense regions listed in
  Table~\ref{tab:extdmassdens}. The predicted fluxes assume a CR
  enhancement factor determined in~\citet{hess_w28}, except for HESS
  J1800-240\,C, $k_{{\rm CR}}$ has been calculated here. We treat the
  flux from the NE shocked cloud as an upper limit, as the estimated
  mass is an upper limit.
  \label{tab:highenergy}}
\begin{tabular}{lcc}
\hline
Region	&	$k_{{\rm CR}}$	&	Expected Flux $F(E \geq 1\,{\rm TeV})$	\\
& & [ph\,\cmsqr \,s$^{-1}$] \\
\hline					
NE Shocked Cloud	&	13	        &  $<$\,5.1$\times10^{-13}$ \\
G5.89-0.39 \hii Region	&	18         	&	7.0$\times10^{-13}$ \\
G5.89-0.39 NW Arm	&	18	        &	5.4$\times10^{-14}$ \\
G5.89-0.39 SE Ridge	&	18	        &	1.6$\times10^{-13}$ \\
G6.1-0.6 Region	        &	14$^{*}$	&	6.5$\times10^{-14}$ \\
G6.225-0.569 Region	&	14$^{*}$	&	1.6$\times10^{-14}$ \\
HESS J1800-240\,C	&	35$^{\dagger}$	&	1.9$\times10^{-15}$ \\
\hline
\multicolumn{3}{l}{\small $^{*}$ $k_{{\rm CR}}$ from~\citet{hess_w28} has been re-scaled}\\
\multicolumn{3}{l}{\small assuming a $d=2$\,kpc.}\\
\multicolumn{3}{l}{\small $^{\dagger}$ $k_{{\rm CR}}$ calculated here using the CO\,(1-0) emission from} \\
\multicolumn{3}{l}{\small ~\citet{nanten10} }\\
\multicolumn{3}{l}{\small $^{\ddagger}$ Assuming the compact-source size mass indicated in Table~\ref{tab:massdens}}\\
\hline
\end{tabular}
\end{table}

\section{Summary and Conclusions}
\label{sec:conc}

Using the Mopra 22\,m telescope, we have conducted 7\,mm molecular
line mapping covering dense and disrupted gas toward the W28 SNR
field. We have followed up our previous 12\,mm line study of the
molecular clouds towards the TeV gamma-ray peaks observed by the
H.E.S.S.  telescope, with higher resolution targeted
observations towards the NE cloud/shock interaction, HESS J1801-233
and the bright and energetic UC-\hii regions in the southern cloud
HESS J1800-240\,B.

Sites of shocks, outflows and disruption are revealed with SiO. In the
majority of cases, SiO detections are towards sites of stellar
activity and \hii regions, however, SiO is also detected in the NE
cloud/SNR shock interaction region with broad line
profiles. Interestingly, in this shock region the SiO emission is
bounded by clusters of 1720\,OH masers, indicating the downstream 
direction of the shock is towards W28.  Other lines detected are
44\,GHz CH$_{3}$\,OH masers and cyanopolyynes HC$_{n}$N ($n=3,5,7$)
which are again preferentially associated with star formation and \hii
regions.

The CS cores are typically found towards CO peaks and trace conditions
ranging from quiescent cloud cores to \hii regions. The obvious
exception is the NE cloud/SNR shock interaction region, which exhibits
broad line emission FWHM\,$>10$\,\kms in all detected lines.

Based on CS isotopologue ratios from our position-switched deep
pointing observations, we are able to estimate the upper state CS
column density. Temperature estimates from our previous \nh3
observations allow the LTE molecular hydrogen mass and density to be
estimated. Our results presented in Table~\ref{tab:massdens} assume a
compact-source core radius $r=0.2$\,pc, however, scaling factors for
various core radii are also included.\\
We have estimated the mass of the extended dense NE cloud and several
of the dense clumps in the southern clouds, assuming the CS
isotopologue emission is contained within a prolate ellipsoid.  Under
these assumptions, we estimate there to be 5.6$\times10^{4}$\,\msun,
with H$_{2}$ number density 6.1$\times10^{4}$\,\cmcube {\bf (as an upper limit)} contained
towards the NE HESS J1801-233 TeV source. We find that
$\sim4\times10^{49}$\,erg of kinetic energy is required to produce the
broad emission in the CS line profile towards the NE cloud, which is a
few percent of the typical $10^{51}$\,erg of kinetic energy released
in a SNR.

For the southern cloud towards HESS J1800-240\,B
we estimate that $\sim70$\% of the gas mass is contained in dense clumps 
surrounding the \hii regions. The southern clouds show no signs of 
external disruption, providing no evidence that the W28 SNR 
shock has influenced the region.

Although our CS masses are treated as upper limits, we note that 
a somewhat independent estimate of the CS mass for clumps not part
of the NE Shocked Cloud comes from the virial theorem. For these cases
the fact that the virial mass is similar to the LTE mass within a factor 2 to 3 suggests our
choice of CS abundance is adequate for order of magnitude estimates.
For the NE Shocked Cloud the virial theorem will likely not apply.
However, as discussed 
in  section~\ref {sec:NEcloud}, the NE Shocked Cloud mass derived from our CS 
observations is ~10\% larger than the mass from CO observations, 
indicating that our CS abundance may be slightly overestimated here.

Additional systematic uncertainties arise from the unknown spectrum of CRs 
entering the dense clouds, and, penetrating their interiors as a result of 
energy-dependent diffusion. Equation~\ref{eqn:TeVFlux} assumes that 
a CR spectral integral index of $\Gamma =-1.6$ as expected for Galactic CRs 
(GCRs) impacting passive clouds some distance away. The spectral indices
 of the TeV emission towards W28 are in fact found to be similar to GCRs
 so Equation~\ref{eqn:TeVFlux} should be adequate in this regard.

The striking spatial match between the TeV emission and the dense
molecular gas suggests that the CRs are able to penetrate some
distance into the dense cloud core. However, the energy dependent
diffusion of CRs into the cloud cores~\citep{gabici}
could easily suppress the predicted emission for low energies $E<1$ TeV 
below that suggested by Equation~\ref{eqn:TeVFlux}. The level of suppression
is strongly dependent on the largely unknown magnetic field structure and 
turbulence in the clump and is a topic beyond the scope of this paper.

Overall, we would like to emphasise that the uncertainties discussed above
clearly motivate the need for new 1 arcmin angular resolution TeV observations 
(for example by CTA) in order to begin probing the diffusion properties of CR into dense 
molecular clouds.

This work is part of our ongoing study into the molecular gas towards
the W28 region, and follows our earlier 12\,mm line mapping
campaign. Additional observations of the NE cloud SNR interaction
region have been conducted with Mopra at 12\,mm in order to extract
detailed \nh3 inversion spectra across the cloud core. Results of
these observations will be the focus of a future paper to further
understand the molecular gas properties in the interesting NE SNR
shock/molecular cloud interaction region.

\section*{Acknowledgements}
We thank Cormac Purcell for his baseline fitting and removal script.
This work was supported by Australian Research Council grants
(DP0662810, DP1096533). The Mopra Telescope is part of the Australia Telescope
and is funded by the Commonwealth of Australia for operation as a
National Facility managed by CSIRO. The University of New South Wales
Mopra Spectrometer Digital Filter Bank used for these Mopra
observations was provided with support from the Australian Research
Council, together with the University of New South Wales, University
of Sydney, Monash University and the CSIRO.

\clearpage
\appendix
\section{Additional Figures and Tables}

\begin{figure}
\includegraphics[width=\columnwidth]{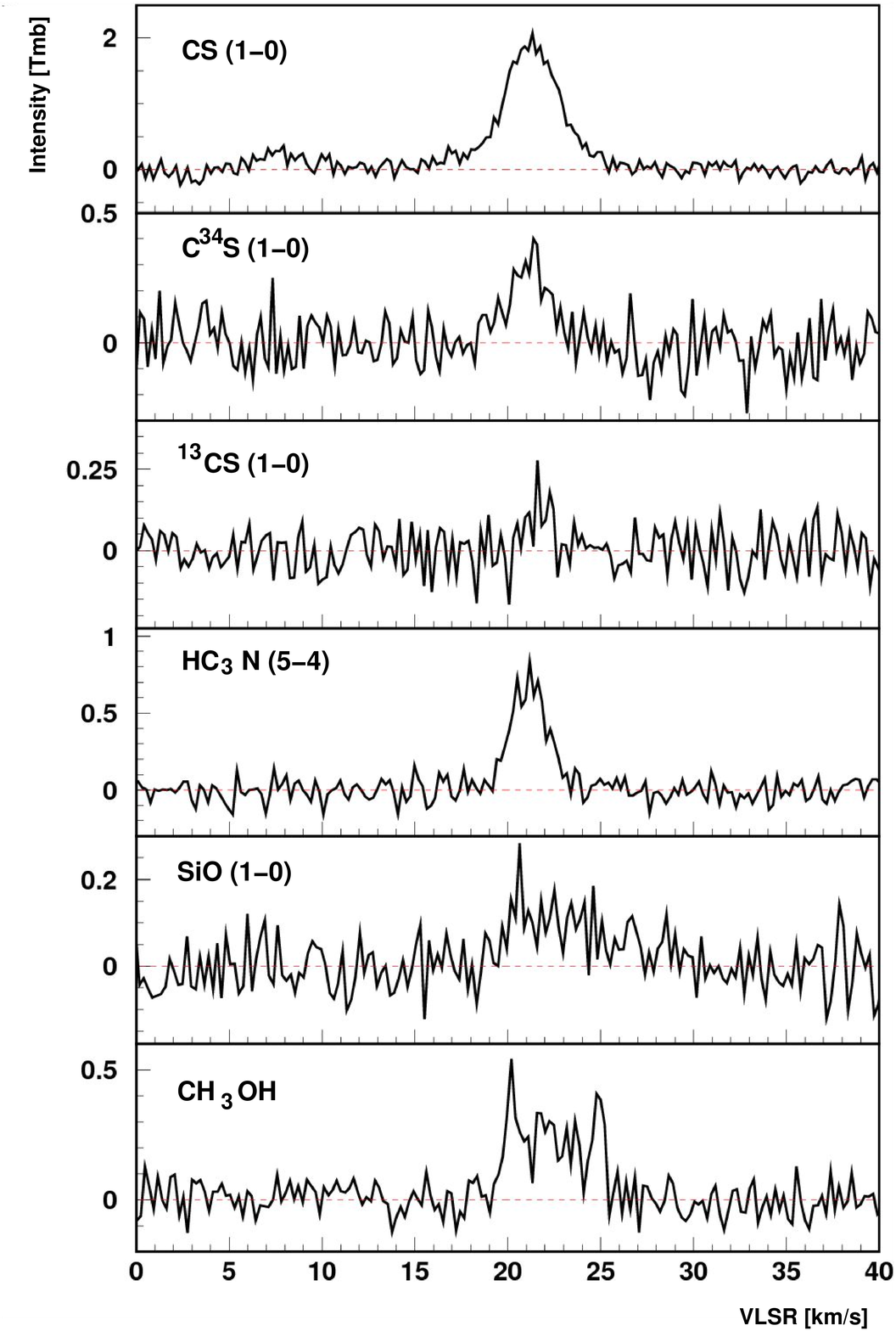}
\caption{Spectra for the line emission detected towards the IRAS
  17589-2312 star formation core from mapping data.}
\label{fig:core1spec}
\end{figure}

\begin{figure}
\centering
\includegraphics[width=\columnwidth]{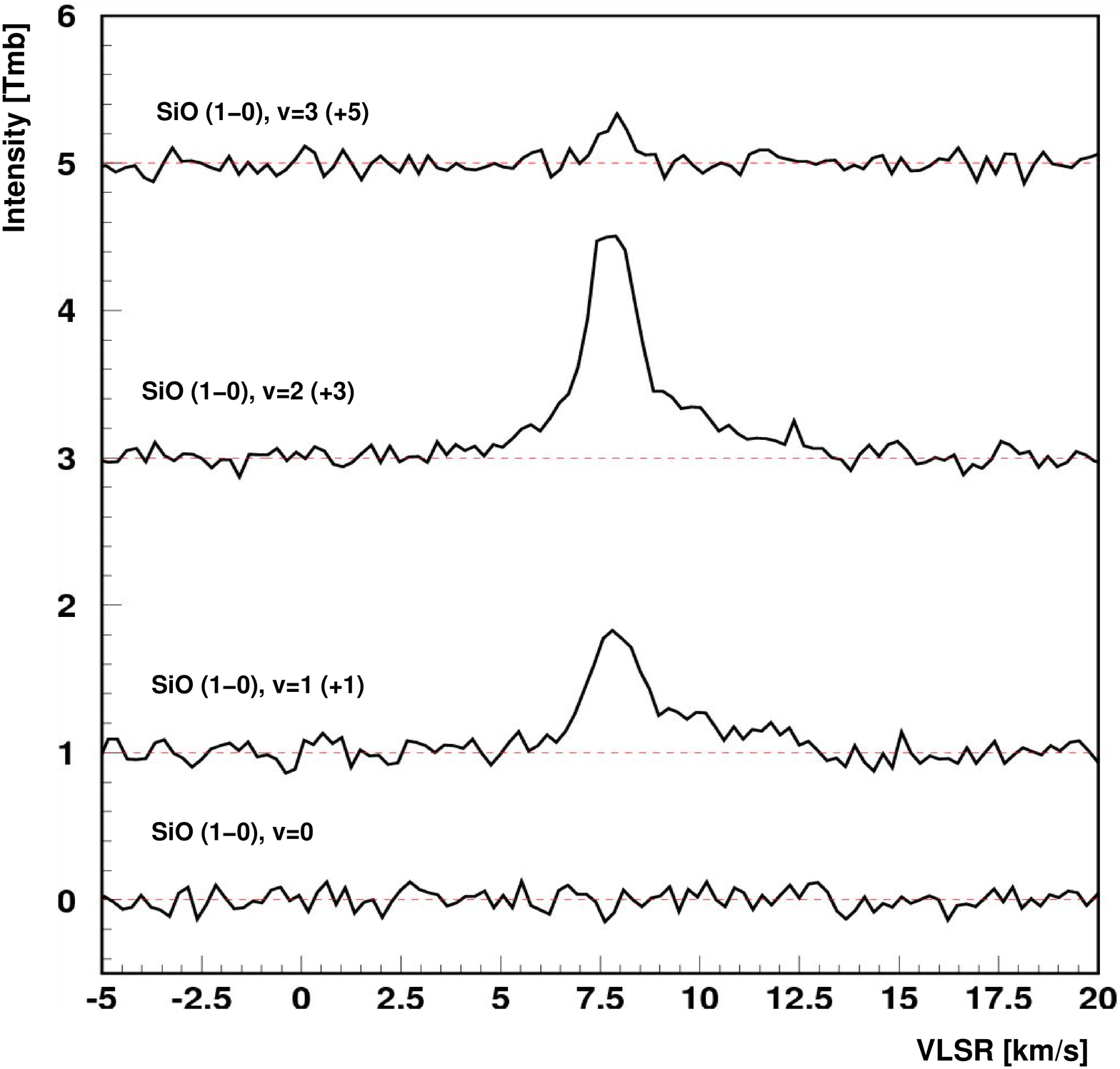}
\caption{Spectra of the SiO\,(1-0) $v=1,2,3$ lines coming from the
  V5357\,Sgr Variable Star of Mira Cet type.}
\label{fig:SiO123}
\end{figure}

\begin{figure}
\includegraphics[width=\columnwidth]{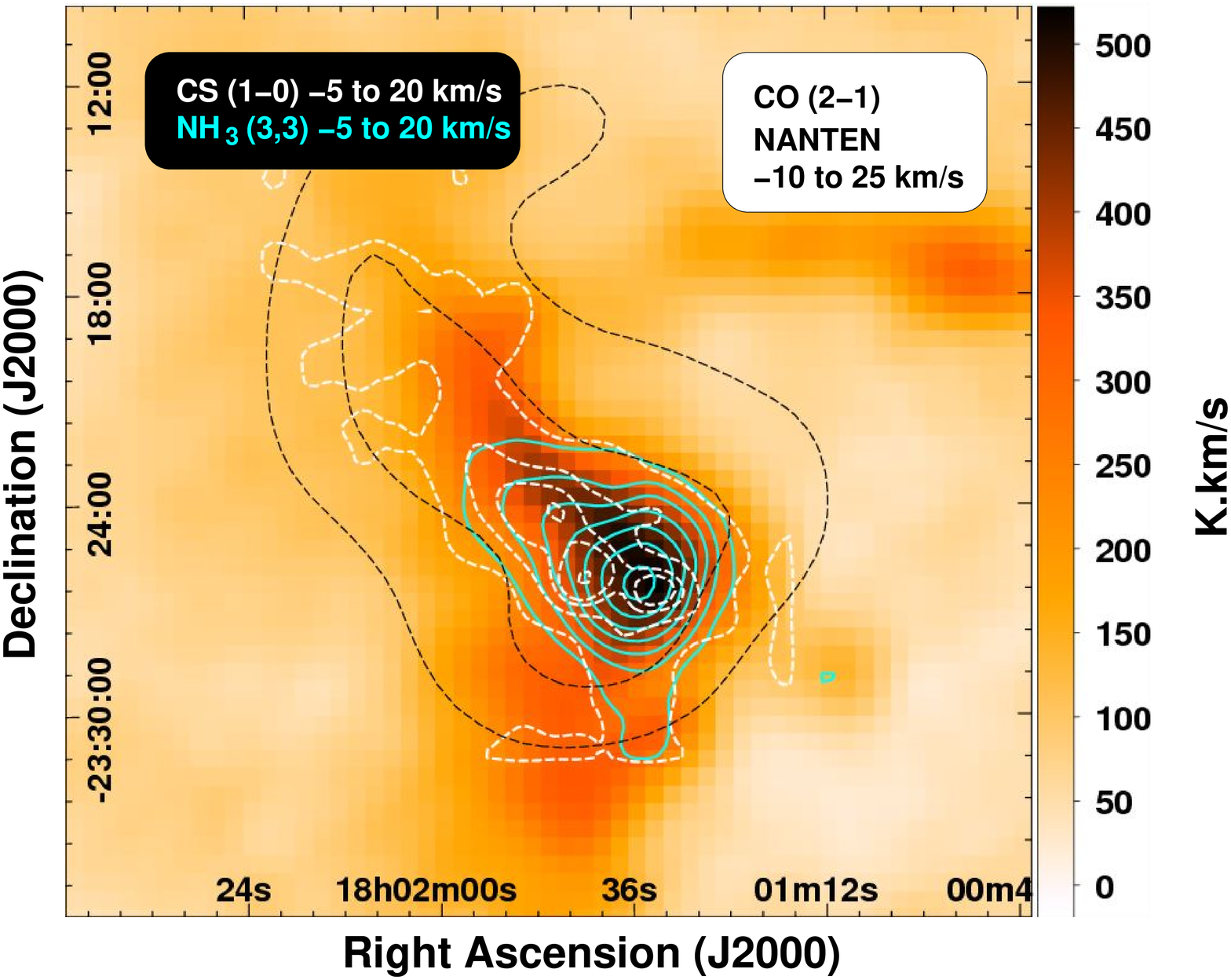}
\caption{Image showing the relationship between the diffuse and dense
  gas in the NE cloud. The colour scale is the NANTEN2 CO\,(2-1) -10 to
  25\,\kms integrated intensity image. The black dased contours are
  the H.E.S.S. TeV emission (4 \& 5\,$\sigma$ levels). The white
  dashed contours are the CS\,(1-0) -5 to 20\,\kms integrated emission
  as seen on Figure~\ref{fig:intmultipanel}. The solid cyan contours
  are the \nh3(3,3) -5 to 20\,\kms integrated emission
  (1,3,5,7,9,11,13\,\Kkms levels) from~\citet{me}.}
\label{fig:densediffuse}
\end{figure}

\clearpage

\begin{table*}
\centering
\caption{Line parameters from a Gaussian fit to the spectra, for lines
  detected in deep pointing observations. \label{app:DPlineFit}}
\begin{tabular}{lcccc}
\hline
Region	&	Peak \tmb			&	$\int$\,\tmb\,$dv$			&	\vlsr			&	FWHM			\\
	&	[$K$]			&	[\Kkms]			&	[\kms]			&	[\kms]			\\
\hline																	
\\
\multicolumn{5}{c}{-- DP-1 --}	\\																
CH$_{3}$OH	&	0.03	$\pm$	0.003	&	0.55	$\pm$	0.09	&	11.70	$\pm$	1.072	&	18.74	$\pm$	2.29	\\
SiO\,(1-0)	&	0.09	$\pm$	0.003	&	2.15	$\pm$	0.12	&	13.12	$\pm$	0.395	&	22.71	$\pm$	0.98	\\
HC$_{3}$N\,(5-4)	&	0.04	$\pm$	0.006	&	0.61	$\pm$	0.14	&	12.18	$\pm$	1.215	&	15.23	$\pm$	2.65	\\
\13CS\,(1-0)	&	0.02	$\pm$	0.003	&	0.58	$\pm$	0.13	&	13.68	$\pm$	1.778	&	25.66	$\pm$	4.63	\\
\C34S\,(1-0)	&	0.07	$\pm$	0.006	&	1.39	$\pm$	0.19	&	12.30	$\pm$	0.603	&	17.80	$\pm$	2.05	\\
CS\,(1-0)	&	1.06	$\pm$	0.005	&	20.46	$\pm$	0.16	&	13.05	$\pm$	0.043	&	18.09	$\pm$	0.11	\\
\\
\multicolumn{5}{c}{-- DP-2 --}	\\																
CH$_{3}$OH	&	0.06	$\pm$	0.004	&	0.98	$\pm$	0.09	&	5.67	$\pm$	0.458	&	15.31	$\pm$	1.08	\\
SiO\,(1-0)	&	0.13	$\pm$	0.004	&	2.29	$\pm$	0.11	&	6.10	$\pm$	0.231	&	15.99	$\pm$	0.58	\\
HC$_{3}$N\,(5-4)	&	0.10	$\pm$	0.007	&	1.16	$\pm$	0.12	&	6.87	$\pm$	0.361	&	10.40	$\pm$	0.84	\\
\13CS\,(1-0)	&	0.07	$\pm$	0.006	&	0.79	$\pm$	0.10	&	8.43	$\pm$	0.352	&	10.49	$\pm$	1.04	\\
\C34S\,(1-0)	&	0.20	$\pm$	0.007	&	1.99	$\pm$	0.12	&	7.55	$\pm$	0.168	&	9.60	$\pm$	0.45	\\
CS\,(1-0)	&	2.37	$\pm$	0.007	&	28.27	$\pm$	0.13	&	7.37	$\pm$	0.015	&	11.20	$\pm$	0.04	\\
\\
\multicolumn{5}{c}{-- DP-3 --}	\\																
CS\,(1-0)	&	1.39	$\pm$	0.064	&	3.28	$\pm$	0.22	&	8.98	$\pm$	0.043	&	2.21	$\pm$	0.11	\\
\\
\multicolumn{5}{c}{-- DP-4 --}	\\																
CH$_{3}$OH	&	3.65	$\pm$	0.014	&	4.46	$\pm$	0.03	&	9.00	$\pm$	0.002	&	1.15	$\pm$	0.01	\\
SiO\,(1-0)	&	0.25	$\pm$	0.006	&	8.98	$\pm$	0.05	&	12.35	$\pm$	0.011	&	19.56	$\pm$	0.02	\\
HC$_{5}$N\,(17-16)	&	0.59	$\pm$	0.005	&	5.15	$\pm$	0.17	&	9.17	$\pm$	0.152	&	3.29	$\pm$	0.53	\\
HC$_{3}$N\,(5-4)	&	4.12	$\pm$	0.009	&	2.07	$\pm$	0.04	&	9.09	$\pm$	0.024	&	3.56	$\pm$	0.05	\\
\13CS\,(1-0)	&	0.40	$\pm$	0.013	&	15.63	$\pm$	0.07	&	9.35	$\pm$	0.005	&	3.64	$\pm$	0.01	\\
HC$_{5}$N\,(16-15)	&	0.52	$\pm$	0.009	&	1.53	$\pm$	0.06	&	8.97	$\pm$	0.037	&	3.10	$\pm$	0.10	\\
\C34S\,(1-0)	&	1.15	$\pm$	0.013	&	1.72	$\pm$	0.06	&	9.29	$\pm$	0.037	&	3.67	$\pm$	0.08	\\
CS\,(1-0)	&	9.50	$\pm$	0.015	&	4.50	$\pm$	0.09	&	9.26	$\pm$	0.022	&	4.01	$\pm$	0.06	\\
\\
\multicolumn{5}{c}{-- DP-5 --}	\\																
\C34S\,(1-0)	&	0.29	$\pm$	0.022	&	0.31	$\pm$	0.04	&	12.97	$\pm$	0.037	&	1.01	$\pm$	0.09	\\
CS\,(1-0)	&	2.80	$\pm$	0.022	&	3.34	$\pm$	0.04	&	13.04	$\pm$	0.004	&	1.12	$\pm$	0.01	\\
\hline																	
\end{tabular}
\end{table*}


\begin{thebibliography}{99}

\bibitem[\protect\citeauthoryear{Abdo \etal}{2010}]{fermi_w28} Abdo A. A., \etal (Fermi Collab.), 2010, ApJ 718, 348
\bibitem[\protect\citeauthoryear{Acciari \etal}{2009}]{ic443_veritas} Acciari V. A., \etal (VERITAS Collab.), 2009, ApJ, 698, L133
\bibitem[\protect\citeauthoryear{Acord \etal}{1997}]{acord} Acord J. M., Walmsley C. M. \& Churchwell E., 1997, ApJ, 475, 693
\bibitem[\protect\citeauthoryear{Aharonian\,}{1991}]{aharonian_1991} Aharonian F. A., 1991, Ap\&SS, 180, 305
\bibitem[\protect\citeauthoryear{Aharonian \etal}{1994}]{aharonian1994} Aharonian F. A., Drury L. O'C. \& V\"{o}lk H. J., 1994, A\&A, 285, 645
\bibitem[\protect\citeauthoryear{Aharonian \etal}{2004}]{hess_1745} Aharonian F., \etal (H.E.S.S. Collab.), 2004, A\&A, 425, L13
\bibitem[\protect\citeauthoryear{Aharonian \etal}{2006}]{hess_CMZ} Aharonian F., \etal (H.E.S.S. Collab.), 2006, Nature, 439, 695
\bibitem[\protect\citeauthoryear{Aharonian \etal}{2008a}]{hess_ctb37a} Aharonian F., \etal (H.E.S.S. Collab.), 2008a, A\&A, 490, 685
\bibitem[\protect\citeauthoryear{Aharonian \etal}{2008b}]{hess_w28} Aharonian F., \etal (H.E.S.S. Collab.), 2008b, A\&A, 481, 401
\bibitem[\protect\citeauthoryear{Albert \etal}{2008}]{ic443_magic} Albert J., \etal (MAGIC Collab.), 2008, ApJ, 674, 1037
\bibitem[\protect\citeauthoryear{Araudo \etal}{2007}]{araudo} Araudo A. T., Romero G. E., Bosch-Ramon V. \& Paredes J. M., 2007, A\&A, 476, 1289
\bibitem[\protect\citeauthoryear{Arikawa \etal}{1999}]{arikawa} Arikawa Y., Tatematsu K., Sekimoto Y. \& Takahashi T., 1999, PASJ, 51, L7
\bibitem[\protect\citeauthoryear{Bronfman \etal}{1996}]{bronfman} Bronfman L., Nyman L.-\AA. \& May J., 1996, A\&ASS, 115, 81
\bibitem[\protect\citeauthoryear{Brogan \etal}{2006}]{brogan} Brogan C. L., Gelfand J. D., Gaensler B. M., Kassim N. E. \& Lazio T. J. W., 2006, ApJ, 639, L25
\bibitem[\protect\citeauthoryear{Crutcher\,}{1991}]{crutcher} Crutcher R. M., 1991, ApJ, 520, 706
\bibitem[\protect\citeauthoryear{Churchwell \etal}{1990}]{churchwell} Churchwell E., Walmsley C. M. \& Cesaroni R., 1990, A\&A~SS, 83, 199
\bibitem[\protect\citeauthoryear{Choi \etal}{1993}]{choi} Choi M., Evans II N. J. \& Jaffe D. T., 1993, 417, 624
\bibitem[\protect\citeauthoryear{Claussen \etal}{1997}]{claussen} Claussen M. J., Frail D. A., Goss W. M. \& Gaume R. A., 1997, ApJ, 489, 143
\bibitem[\protect\citeauthoryear{Codella \etal}{1995}]{codella} Codella C., Palumbo G.G.C., Pareschi G., Scappini F., Caselli P. \& Attol	ini M. R., 1995, MNRAS, 276, 57
\bibitem[\protect\citeauthoryear{The CTA Consortium\,}{2010}]{CTAdesign} The CTA Consortium, 2010, {\it arXiv:1008.3703v2}
\bibitem[\protect\citeauthoryear{Dubner \etal}{2000}]{dubner} Dubner G. M., Vel\'azquez P. F., Goss W. M., Holdaway M. A., 2000, AJ, 120, 1933
\bibitem[\protect\citeauthoryear{Feldt \etal}{2003}]{feldt} Feldt M., Puga E., Lenzen R., Henning TH., Brandner W., Stecklum B., Lagrange A.-M., Gendron E. \& Rousset G., 2003, ApJ, 599, L91
\bibitem[\protect\citeauthoryear{Feinstein \etal}{2009}]{feinstein} Feinstein F., Fiasson A., Gallant Y., Chaves R. C. G., Marandon V., de Naurois M., Kosack K. \& Rowell G. (for H.E.S.S. Collab.), 2009, AIP Conf. Proc. 1112, 54
\bibitem[\protect\citeauthoryear{Fontani \etal}{2010}]{fontani} Fontani F., Cesaroni R. \& Furuya R.S, 2010, A\&A 517, A56 
\bibitem[\protect\citeauthoryear{Frail \etal}{1994}]{frail} Frail D. A., Goss W. M. \& Slysh V. I., 1994, ApJ, 424, L111
\bibitem[\protect\citeauthoryear{Frerking \etal}{1980}]{frerking} Frerking M. A., Wilson R. W. \& Linke R. A., 1980, ApJ, 240, 65
\bibitem[\protect\citeauthoryear{Fujita \etal}{2009}]{fujita2009} Fujita Y., Ohira Y., Tanaka S. J. \& Takahara F. 2009, ApJ, 707, L179
\bibitem[\protect\citeauthoryear{Fukui \etal}{2008}]{nanten21} Fukui Y. \etal (NANTEN Collab.), 2008, AIP Conf. Proc., 1085, 104
\bibitem[\protect\citeauthoryear{Gabici \etal}{2007}]{gabici} Gabici S., Aharonian F. A. \& Blasi P. 2007, Astrophys. Space Sci., 309, 365
\bibitem[\protect\citeauthoryear{Gabici \etal}{2010}]{gabici2010} Gabici S., Casanova S., Aharonian F. A. \& Rowell G, 2010, in Proc. Soci\'et\'e Francaise d'Astronomie et 
d'Astrophysique, 237 ({\it arXiv:1007.4869v1})
\bibitem[\protect\citeauthoryear{Goudis\,}{1976}]{goudis} Goudis C., 1976, Ap\&SS, 40, 91
\bibitem[\protect\citeauthoryear{Giuliani \etal}{2010}]{agile} Giuliani A., \etal (AGILE Collab.), 2010, A\&A 516, L11
\bibitem[\protect\citeauthoryear{Goldsmith \& Langer\,}{1999}]{goldsmith_langer} Goldsmith P. F. \& Langer W. D., 1999, ApJ, 517, 209
\bibitem[\protect\citeauthoryear{Gomez\,}{1991}]{gomez} G\'{o}mez Y., Rodr\'{i}guez L. F., Garay G., Moran J. M., 1991, ApJ, 377, 519
\bibitem[\protect\citeauthoryear{Gusdorf \etal}{2008a}]{Gusdorf:2008a} Gusdorf A., Cabrit S., Flower D. R. \& Pineau Des For\^{e}ts G., 2008a, A\&A, 482, 809
\bibitem[\protect\citeauthoryear{Gusdorf \etal}{2008b}]{Gusdorf:2008b} Gusdorf A., Pineau des For\^{e}ts G., Cabrit S \& Flower D. R., 2008b, A\&A, 490, 695
\bibitem[\protect\citeauthoryear{Harvey \& Forveille\,}{1988}]{harvey} Harvey P. M. \& Forveille T., 1988, A\&A, 197, L19
\bibitem[\protect\citeauthoryear{Hewitt \& Yusef-Zadeh\,}{2009}]{hewitt} Hewitt J. W., Yusef-Zadeh F., 2009, ApJ, 694, L16
\bibitem[\protect\citeauthoryear{Hunter \etal}{2008}]{hunter} Hunter T.L., Brogan C.L., Indebetouw R., Cyganowski C.J., 2008, ApJ, 680, 1271
\bibitem[\protect\citeauthoryear{Inoue \etal}{2010}]{inoue} Inoue T., Yamazaki R. \& Inutsuka S., 2010, ApJ, 723, L108.
\bibitem[\protect\citeauthoryear{Irvine \etal}{1987}]{irvine} Irvine W. M., Goldsmith P. F. \& Hjalmarson A., 1987, Chemical Abundances in Molecular Clouds. In Hollenbach D. J., Thronson Jr. H. A. (editors) Interstellar Processes, 1987, Reidel Deordrecht
\bibitem[\protect\citeauthoryear{Kaspi \etal}{1993}]{kaspi} Kaspi V. M., Lyne A. G., Manchester R. N., Johnston S., D'Amico N. \& Shemar S. L., 1993, ApJ, 409, L57
\bibitem[\protect\citeauthoryear{Kim \& Koo\,}{2001}]{kimkoo2001} Kim K. \& Koo B., 2001, ApJ, 549, 979
\bibitem[\protect\citeauthoryear{Kim \& Koo\,}{2003}]{kimkoo2003} Kim K. \& Koo B., 2003, ApJ, 596, 362
\bibitem[\protect\citeauthoryear{Kuchar \& Clark\,}{1997}]{kuchar} Kuchar T. A., Clark F. O., 1997, ApJ, 488, 224
\bibitem[\protect\citeauthoryear{Lefloch \etal}{2008}]{lefloch} Lefloch B., Cernicharo J., Pardo J. R., 2008, A\&A, 489, 157
\bibitem[\protect\citeauthoryear{Li \& Chen}{2010}]{li2010} Li H. \& Chen Y. 2010, MNRAS, 409, L35
\bibitem[\protect\citeauthoryear{Linke \& Goldsmith\,}{1980}]{linke} Linke R. A. \& Goldsmith P. F., 1980, ApJ, 235, 437
\bibitem[\protect\citeauthoryear{Liszt\,}{2009}]{liszt} Liszt H.S., 2009, A\&A, 508, 1331
\bibitem[\protect\citeauthoryear{Lockman\,}{1989}]{lockman} Lockman F. J., 1989, ApJSS, 71, 469
\bibitem[\protect\citeauthoryear{Lozinskaya\,}{1981}]{lozinskaya} Lozinskaya T. A., 1981, Sov. Astron. Lett., 7, 17
\bibitem[\protect\citeauthoryear{Martin-Pintado \etal}{2000}]{Martin-Pintado:2000} Martin-Pintado J., de\,Vincente P., Rodr\'{i}guez-Fern\'{a}ndez N. J., Fuente A. \& Planesas P., 2000, A\&A, 356, L5
\bibitem[\protect\citeauthoryear{Motogi \etal}{2010}]{motogi} Motogi K., Sorai K., Habe A., Honma M., Kobayashi H. \& Sato K., 2011, PASJ, 63, 31
\bibitem[\protect\citeauthoryear{Mizuno \& Fukui\,}{2004}]{nanten10} Mizuno A. \& Fukui Y., 2004, ASP Conf. Proc., 317, 59
\bibitem[\protect\citeauthoryear{Nicholas \etal}{2010}]{me} Nicholas B., Rowell G., Burton M.G., Walsh A., Fukui Y., Kawamura A., Longmore S. \& Keto E., 2011, MNRAS, 411, 1367
\bibitem[\protect\citeauthoryear{Ohira \etal}{2011}]{ohira_2010} Ohira Y., Murase K. \& Yamazaki R., 2011, MNRAS 410, 1577
\bibitem[\protect\citeauthoryear{Reach \etal}{2005}]{reach} Reach W. T., Rho J., Jarrett T. H., 2005, ApJ 618, 297
\bibitem[\protect\citeauthoryear{Rho \& Borkowski\,}{2002}]{rho2002} Rho J. \& Borkowski K. J., 2002, ApJ, 575, 201
\bibitem[\protect\citeauthoryear{Schilke \etal}{1997}]{Schilke:1996} Schilke P., Walmsley C. M., Pineau Des For\^{e}ts G. \& Flower D. R., 1997, A\&A, 321, 293
\bibitem[\protect\citeauthoryear{Seta \etal}{1998}]{seta_1998} Seta M., Hasegawa T., Dame T. M., Sakamoto S., Oka T., Handa T., Hayashi M., Morino J., Sorai K. \& Usuda K. S., 1998, ApJ, 505, 286
\bibitem[\protect\citeauthoryear{Sollins \etal}{2004}]{sollins} Sollins P. K., Hunter T. R., Battat J., Beuther H., Ho P. T. P., Lim J., Liu S. J., Ohashi N., \etal, 2004, ApJ, 616, L35
\bibitem[\protect\citeauthoryear{Thompson \& Macdonald\,}{1999}]{thompson} Thompson M. A. \& Macdonald G. H., 1999, A\&ASS, 135, 531
\bibitem[\protect\citeauthoryear{Torres \etal}{2003}]{torres} Torres D. F., Romero G. E., Dame T. M., Combi J. A. \& Butt Y. M., 2003, Phys. Rep. 382, 303
\bibitem[\protect\citeauthoryear{Uchiyama \etal}{2010}]{uchiyama} Uchiyama Y., Blandford R. D., Funk S., Tajima H. \& Tanaka T., 2010, ApJ, 723, L122
\bibitem[\protect\citeauthoryear{Ueno \etal}{2003}]{ueno2003} Ueno M., Bamba A. \& Koyama K., 2003, proc. 28$^{\rm th}$ ICRC, 2401
\bibitem[\protect\citeauthoryear{Urquhart \etal}{2010}]{mopra_beam} Urquhart J. S., Hoare M. G., Purcell C. R., Brooks K. J., Voronkov M. A., Indermuehle B. T., Burton M. G., Tothill N. F. H. \& Edwards P. G., 2010, PASA, 27, 321 
\bibitem[\protect\citeauthoryear{Vel\'azquez \etal}{2010}]{velazquez} Vel\'azquez P. F., Dubner G. M., Goss W. M., Green A. J., 2002, AJ, 124, 2145
\bibitem[\protect\citeauthoryear{Voronkov \etal}{2010}]{voronkov} Voronkov M. A., Caswell J. L., Ellingsen S. P. \& Sobolev A. M., 2010, MNRAS, 405, 2471
\bibitem[\protect\citeauthoryear{Wootten\,}{1981}]{wootten} Wootten A., 1981, ApJ, 245, 105
\bibitem[\protect\citeauthoryear{Yamazaki \etal}{2006}]{yamazaki2006} Yamazaki R., Kohri K., Bamba A., Yoshida T., Tsuribe T. \& Takahara F., 2006, MNRAS, 371, 1975
\bibitem[\protect\citeauthoryear{Zhou \etal}{1989}]{zhou} Zhou S., Wu Y., Evans II N. J., Fuller G. A. \& Myers P. C., 1989, ApJ, 346, 168
\bibitem[\protect\citeauthoryear{Zijlstra \etal}{1990}]{zijlstra} Zijlstra A. A, Pottasch S. R, Engels D., Roelfsema P. R., Hekkert P. T. L. \& Umana G., 1990, MNRAS, 246, 217
\bibitem[\protect\citeauthoryear{Zinchenko \etal}{1994}]{zinchenko} Zinchenko I., Forsstr\"om V., Lapinov A. \& Mattila K., 1994, A\&A, 288, 601


\end{thebibliography}
\end{document}